\documentclass[11pt,letterpaper]{JHEP3}

\usepackage{epsfig,multicol}
\usepackage{amsmath}
\usepackage{amssymb}
\usepackage{bbm}
\usepackage{yfonts}

\newcommand{\labell}[1]{\label{#1}}  

\newcommand{\be}{\begin{equation}}
\newcommand{\ee}{\end{equation}}
\newcommand{\bea}{\begin{eqnarray}}
\newcommand{\eea}{\end{eqnarray}}
\newcommand{\ba}{\begin{eqnarray}}
\newcommand{\ea}{\end{eqnarray}}

\newcommand{\beq}{\begin{equation}}
\newcommand{\eeq}{\end{equation}}
\newcommand{\beqa}{\begin{eqnarray}}
\newcommand{\eeqa}{\end{eqnarray}}
\newcommand{\beqar}{\begin{eqnarray*}}
\newcommand{\eeqar}{\end{eqnarray*}}

\newcommand{\reef}[1]{(\ref{#1})}

\newcommand{\eg}{{\it e.g.,}\ }
\newcommand{\ie}{{\it i.e.,}\ }

\newcommand{\mt}[1]{\textrm{\tiny #1}}

\newcommand{\veps}{\varepsilon}
\newcommand{\X}{\mathcal{X}}
\newcommand{\W}{\mathcal{W}}
\newcommand{\Z}{\mathcal{Z}}
\newcommand{\A}{\mathcal{A}}
\newcommand{\B}{\mathcal{B}}
\newcommand{\C}{\mathcal{C}}

\newcommand{\E}{\mathcal{E}}

\newcommand{\nvec}{{\mbf n}} 
\newcommand{\tL}{\tilde{L}}

\renewcommand{\eqref}[1]{(\ref{#1})}

\newcommand{\la}{\lambda}
\newcommand{\lp}{\ell_{\mt P}}
\newcommand{\mbf}{\mathbf}
\newcommand{\fin}{f_\infty}
\newcommand{\ff}{f_{\infty}}

\newcommand{\scrw}{\textswab{w}}
\newcommand{\scrq}{\textswab{q}}

\title{Holographic studies of quasi-topological gravity}

\author{Robert C. Myers,$^{a}$ Miguel F. Paulos$^{b}$ and Aninda Sinha$^a$ \\
$^a$ {\it Perimeter Institute for Theoretical Physics, Waterloo,
Ontario N2L 2Y5, Canada}\\
$^b$ {\it Department of Applied Mathematics and Theoretical
    Physics, Cambridge CB3 0WA, UK}\\

\vskip .5cm

{\rm E-mail:}\ \ {\tt rmyers,$\,$asinha@perimeterinstitute.ca, \
m.f.paulos@damtp.cam.ac.uk}}

\abstract{Quasi-topological gravity is a new gravitational theory including
curvature-cubed interactions and for which exact black hole solutions
were constructed. In a holographic framework, classical
quasi-topological gravity can be thought to be dual to the large $N_c$
limit of some non-supersymmetric but conformal gauge theory. We
establish various elements of the AdS/CFT dictionary for this duality.
This allows us to infer physical constraints on the couplings in the
gravitational theory. Further we use holography to investigate
hydrodynamic aspects of the dual gauge theory. In particular, we find
that the minimum value of the shear-viscosity-to-entropy-density ratio
for this model is $\eta/s \simeq 0.4140/(4\pi)$.}


\preprint{arXiv:1004.2055 [hep-th]\\DAMTP-2010-28}

\begin{document}{\vskip 1cm}

\section{Introduction}

The AdS/CFT correspondence has proven a fertile ground for
investigating the properties of strongly coupled gauge theories
\cite{juan,adscft}, in particular the thermodynamic and hydrodynamic
properties of these gauge theories at finite temperature
\cite{hydro,sonrev}. However, such investigations face acute
limitations because at present, we have an insufficient understanding
of string theory in interesting holographic backgrounds, \ie in
spacetimes with Ramond-Ramond fields. Hence the examination of
holographic gauge theories is primarily confined to both the limit of a
large `t Hooft coupling $\lambda$ and a large number of colours $N_c$
where the dual gravitational theory corresponds to (semi-)classical
Einstein gravity with a two-derivative bulk action. However, it is
understood that accounting for higher curvature interactions, or higher
derivative interactions more generally, within a perturbative framework
allows one to begin to consider finite $\lambda$ and finite $N_c$
corrections \cite{BuchelDependence,buchel2,kp,etas,univers,fundeta}. An
alternative point of view would be that admitting such higher curvature
(or higher derivative) interactions introduces new couplings amongst
the operators in the dual CFT, thereby broadening the universality
class of dual CFT's which one can study with holography
\cite{hofmal,hofman,fundeta}. If one examines a point in the space of
CFT's where these new couplings are finite, the higher curvature terms
will now make finite contributions in the analysis of the dual gravity
theory. However, if any higher curvature term were to become important,
the normal expectation is that an infinite number of such terms will
become important at the same time as the background curvature must have
reached the string or Planck scales. The relevance of all these terms
is really signalling that one has entered a regime where the dual
gravitational theory cannot be described by a local quantum field
theory. Hence to properly investigate the effects of these finite CFT
couplings, one is brought back to the question of understanding string
theory in interesting holographic backgrounds.

However, a traditional avenue to progress in theoretical physics is the
study of simplified or toy models which might provide insight into the
behaviour of some complex physical system of interest. Recent work with
Gauss-Bonnet (GB) gravity showed that the utility of such toy models in
a holographic framework \cite{shenker2,shenker1,alex0,spain1,GBanyd,
GBrefs}. In this case, the usual Einstein action is supplemented by a
certain curvature-squared interaction, which corresponds precisely to
the four-dimensional Euler density. With this extension of the usual
Einstein action in the five-dimensional bulk gravity theory, the class
of holographic models is extended to allow independent values of the
two central charges $a$ and $c$ of the dual CFT \cite{duff,two}.
Further it was found that GB gravity still captures certain fundamental
constraints which can also be inferred from direct considerations of
CFT's alone. In particular, consistency of the CFT constrains the
central charges to obey \cite{hofmal}: $1/2\le a/c \le 3/2$. Hence GB
gravity (or more generally Lovelock gravity in higher dimensions
\cite{LLrefs}) provides an interesting toy model to examine questions
related to holographic hydrodynamics, or perhaps the holographic
$c$-theorem \cite{see,msanom}.

Motivated by the success of holographic studies of GB gravity, this
holographic model was recently extended with the introduction of a new
curvature-cubed interaction in quasi-topological gravity \cite{quasi}.
The progress with GB gravity relies on the fact that even though this
is a higher curvature theory of gravity, the holographic calculations
in this model are still under control. This control stems from two
properties of the theory: the equations of motion are only second order
in derivatives and exact black hole solutions have been constructed. In
quasi-topological gravity, exact black hole solutions are again readily
constructed but on general backgrounds the equations of motion will be
fourth order in derivatives \cite{quasi}. Remarkably, however, the
linearized equations of motion in an AdS$_5$ background are again
second order and in fact, match precisely the linearized equations of
the Einstein theory \cite{quasi}. As we will show in the following,
these properties are sufficient to allow us to examine many interesting
features of the holographic framework established by this new toy
model. The new curvature-cubed interactions again expand the class of
CFT's which can be realized with this model. In particular, the new
couplings are generalized such that the dual CFT will not be
supersymmetric \cite{hofmal} and so this holographic model may provide
new insights on non-supersymmetric gauge theories with a conformal
fixed point.

One aspect which we examine with this new holographic model are the
hydrodynamic transport coefficients of the dual CFT, in particular the
shear viscosity. It has been observed that the ratio of shear viscosity
to density entropy of typical holographic fluids is extremely small in
comparison to typical fluids for ordinary matter \cite{hydro}.
Originally it was conjectured that these holographic calculations
provided a universal lower bound, namely $\eta/s\ge1/4\pi$. One piece
of circumstantial evidence for this conjecture came from string theory
were it was found that the leading finite $\lambda$ corrections always
raised $\eta/s$ above the bound for supersymmetric plasmas where $c=a$
\cite{BuchelDependence,buchel2,etas,univers}. However, it is now
understood this KSS bound can be violated in string theory duals of
plasmas where $c\neq a$ by the effect of new curvature-squared
interactions in the gravitational action \cite{kp,fundeta}. However,
the string theory constructions where these higher curvature terms are
under control only allow for small perturbative violations of the KSS
bound. General arguments still suggest that the ratio of the shear
viscosity to entropy density should satisfy some lower bound
\cite{hydro,bek} and so the question naturally arises as to the precise
nature of such a bound. Hence it is certainly of interest to explore
situations where finite violations of the bound occur and the toy
models above provide a framework for such explorations. In particular,
it would be interesting if one was able to show that $\eta/s$ could be
pushed to zero without producing any other pathologies developing in
the theory. With quasi-topological gravity, we find that $\eta/s$
reaches a non-zero lower value in a particular corner of the allowed
space of gravitational couplings.

An outline of the rest of the paper is as follows: In section
\ref{qtgrav}, we present a brief review of quasi-topological gravity in
five dimensions and the black hole solutions of the theory. We begin to
establish the AdS/CFT dictionary for this gravitational theory in
section \ref{central} with a calculation of the central charges of the
dual CFT.  In section \ref{flux}, we adapt the scattering experiments
in the CFT of \cite{hofmal} to a holographic calculation. These
computations yield directly the flux coefficients $t_2$ and $t_4$, but
combined with the expressions for the central charges, we are also able
to express the coefficients $\A$, $\B$ and $\C$, which determine the
three-point functions of the stress tensor, in terms of the
gravitational couplings. Next in section \ref{harmsway}, we consider
various constraints on the gravitational couplings which are required
to ensure the physical consistency of the dual CFT. We consider three
independent constraints: positivity of the central charge $c$,
positivity of the energy fluxes in section \ref{flux} and avoiding
violations of causality. In section \ref{shear}, we examine the
hydrodynamic behaviour of the CFT plasma. In particular, we find that
the minimum value of the ratio of the shear viscosity to the entropy
density in this model is $\eta/s \simeq 0.4140/(4\pi)$. We provide a
preliminary analysis of possible instabilities of the black holes, or
alternatively of a uniform plasma in the dual CFT at finite temperature
in section \ref{plastab}. We conclude with a brief discussion of our
results and future directions in section \ref{discuss}.

\section{Review of quasi-topological gravity} \label{qtgrav}

We begin with a review of some salient features of quasi-topological
gravity. We focus on the five-dimensional version of the gravity
theory, which would be dual to a four-dimensional CFT. The bulk gravity
action can be written as \cite{quasi}:
 \be
I=\frac {1}{2\lp^3}\int d^5x \sqrt{-g}
\left[\frac{12}{L^2}+R+\frac{\lambda}2 L^2 \X_4+\frac 78 \mu L^4
\Z'_5\right]
 \labell{action}
 \ee
where $\X_4$ is the four-dimensional Euler density, as used in GB
gravity
 \beq
 \X_4=R_{\mu\nu\rho\sigma}R^{\mu\nu\rho\sigma}-4\, R_{\mu\nu}R^{\mu\nu}+R^2\ ,
 \labell{euler4}
 \eeq
and $\Z'_5$ is the new curvature-cubed interaction
 \beqa
 \Z'_5&=&
R_{\mu\nu}{}^{\rho\sigma} R_{\rho\sigma}{}^{\alpha\beta}
R_{\alpha\beta}{}^{\mu\nu} +
\frac{1}{14}\left(21\,R_{\mu\nu\rho\sigma}R^{\mu\nu\rho\sigma}
R-120\,R_{\mu\nu\rho\sigma}R^{\mu\nu\rho}{}_{\alpha}R^{\sigma\alpha}
 \right. \labell{result5a}\\
&&\qquad\left.+ 144\,R_{\mu\nu\rho\sigma} R^{\mu\rho}R^{\nu\sigma}
+128\,R_\mu{}^{\nu}R_\nu{}^{\rho}R_\rho{}^{\mu} - 108\,
R_\mu{}^{\nu}R_\nu{}^{\mu}R +11\,R^3\right)\ .\nonumber
 \eeqa
The AdS vacua of this theory have a curvature scale given by
 \be
\frac{1}{\tilde{L}^2} = \frac{\fin}{L^2} \labell{curf}
 \ee
where the constant $\fin$ is determined as one of the roots of
 \be
1-\fin+\lambda\fin^2+\mu\fin^3=0\ . \labell{fin}
 \ee
Note for any choice of the couplings $\la$ and $\mu$, there is at most
one ghost-free AdS vacuum which supports nonsingular black hole
solutions, as described in detail in \cite{quasi}. The solutions
describing planar AdS black holes take the form
 \beq
\mathrm{d}s^2 = \frac{r^2}{L^2}\left(- \frac{f(r)}{\fin}
\,\mathrm{d}t^2 +\mathrm{d}x_1^2+\mathrm{d}x_2^2+\mathrm{d}x_3^2\right)
+\frac{L^2}{r^2 f(r)}\, \mathrm{d}r^2\,, \labell{metric0}
 \eeq
where $f(r)$ is determined by roots of the following cubic equation:
 \be
1-f(r)+\lambda f(r)^2+\mu f(r)^3=\frac{r_0^4}{r^4}\ .\labell{constr1}
 \eeq
For the relevant solutions, the black hole horizon occurs at $r=r_0$,
which is easily seen to yield $f(r=r_0)=0$ as a solution of the above
equation. The Hawking temperature is given by
 \beq
T = \frac{r_0}{\pi L^2 }\frac{1}{\fin^{1/2}} \ . \labell{planarTh}
 \eeq
The energy and entropy densities are simply calculated as \cite{quasi}:
 \beq
\rho = \frac{3r_0^4}{2\lp^3 L^5 \fin^{1/2}}\ , \qquad\qquad
s = \frac{2\pi r_0^3}{\lp^3L^3}\ .\labell{entropy}
 \ee
Further note that these relations satisfy $\rho=\frac{3}{4} T s$, as
expected for a four-dimensional CFT (in the absence of a chemical
potential).

Apart from finding exact black hole solutions, another remarkable
property of quasi-topological gravity is that the linearized graviton
equations in the five-dimensional AdS vacuum take the form
\cite{quasi}:
 \beqa
&&-\frac{1}{2}\left(1-2\lambda f_\infty-3\mu f_\infty^2
\right)\left[\nabla^2h_{a b} + \nabla_a\nabla_b\, h_c{}^c-
\nabla_a\nabla^ch_{cb}- \nabla_b\nabla^ch_{ca}
 \vphantom{\frac{2}{\tilde{L}^2}} \right.
 \labell{fullEin}\\
&&\qquad\qquad\left.-g^\mt{[0]}_{ab}\left(\nabla^2
h_c{}^c-\nabla^c\,\nabla^dh_{cd}\right) +\frac{2}{\tilde{L}^2}\, h_{a
b}-\frac{2}{\tilde{L}^2}\,g^\mt{[0]}_{ab}\,
h_c{}^c\right]=\lp^3\,\hat{T}_{ab}\,. \nonumber
 \eeqa
Here $g^\mt{[0]}_{ab}$ is the background AdS$_5$ metric and $\tilde{L}$
is the curvature scale in eq.~\reef{curf}. Hence the observation is
that in the AdS$_5$ background, the linearized equations are only
second order in derivatives. In fact, up to an overall factor, the
above equations \reef{fullEin} are precisely the same as the linearized
equations for Einstein gravity in an AdS$_5$ background
--- for example, see \cite{standard,rush}. This result contrasts with that for a generic
$R^3$ action which would yield fourth order equations of
motion.\footnote{Recently, other theories of curvature-cubed gravity
with exceptional properties were identified in \cite{aninda,newer}. In
fact, up to a contribution proportional to the six-dimensional Euler
density, the curvature-cubed interaction constructed in five dimensions
by \cite{newer} is identical to that studied here.} However, the same
occurs here for quasi-topological gravity on a general spacetime
geometry. That is, on general backgrounds, the linearized equations are
fourth order in derivatives for the present theory as well.

\section{Central charges} \label{central}

In this section and the following section, we develop the dictionary
relating the couplings in five-dimensional quasi-topological gravity to
parameters which characterize the dual four-dimensional CFT. Since we
are only dealing with the gravitational sector of the AdS theory, we
are looking to examine the behaviour of the stress energy tensor of the
CFT. Two such parameters are the central charges, $c$ and $a$, of the
CFT. We calculate these through their appearance in the trace anomaly
\cite{duff}, using the now standard holographic approach of
\cite{renorm1}. The central charge $c$ also fixes the coefficient of
the leading singularity in the operator product of the stress tensor
with itself \cite{Dobrev,osborn1,cft4}. Hence as a verification of our
first calculation, we also determine $c$ from examining the two-point
function in section \ref{2ptSec}.

\subsection{Holographic trace anomaly} \label{trace}

The two central charges of a four-dimensional CFT can be defined by the
trace anomaly that arises when the CFT is placed on a curved background
geometry \cite{duff}:
 \be
\langle\, T_{a}{}^{a}\,\rangle= \frac{c}{16\pi^2}\, I_4
-\frac{a}{16\pi^2}\, \X_4\,,\labell{tranomaly}
 \ee
where $\X_4$ is the four-dimensional Euler density, whose structure is
given in eq.~\reef{euler4} (although here, $\X_4$ is evaluated for the
four-dimensional background metric of the CFT), and $I_4$ is the square
of the Weyl tensor, \ie
 \be
I_4= C_{abcd}\, C^{abcd}= R_{abcd}\, R^{abcd}-2 \,R_{ab}R^{ab}
+\frac{1}{3}\, R^2\,,
 \labell{weyl2}
 \ee
In order to compute $c$ and $a$ for the CFT dual to quasi-topological
gravity, we follow the holographic procedure described in
\cite{renorm1}. We should note that modifications to the central
charges from $R^2$ interactions were examined previously in \cite{two}
while perturbative corrections coming from $R^3$ interactions were
considered in \cite{bd}. Efficient methods, \ie `short cuts' to
calculate the holographic trace anomalies for an arbitrary
gravitational action are discussed in \cite{msanom,adam}.

Following \cite{renorm1}, we begin with the Fefferman-Graham expansion
 \be
\mathrm{d}s^2=\frac{\tilde{L}^2}{4 \rho^2} \mathrm{d}\rho^2+ \frac{g_{ab}}{\rho}\, \mathrm{d}x^a
\mathrm{d}x^b\,,
 \ee
with
 \be
g_{a b}=g_{(0) a b}+ \rho\, g_{(1)a b}+\rho^2 g_{(2)a b}+ \cdots \,,
 \ee
where the boundary metric $g_{(0)a b}$ corresponds to the background
geometry of the dual CFT. The next step is to substitute this expansion
of the metric into the gravity action \reef{action}. On-shell $g_{(2)}$
drops out and we are left with an action involving $g_{(0)}$ and
$g_{(1)}$. To extract the conformal anomaly, we focus on terms which
when integrated produce a log divergence. This leads to
\begin{eqnarray}
I&=&{\cal N}\int_\epsilon \frac{d\rho}{\rho} \int d^4 x \sqrt{g_{(0)}} \bigg{[} \bigg{(}
t_1\, R^{(0)^2}+t_2\, R^{(0)}_{\,a b} R^{(0)a b}+t_3\, R^{(0)}_{\,abcd}\,
R^{(0)abcd}\bigg{)}
\labell{holact}\\
      &&\qquad\quad + A\, R^{(0)a b} g_{(1)a b}+B\, R^{(0)}\, {\rm tr}\, g_{(1)} +
      C\, {\rm tr}\, g_{(1)}^2+D\, ({\rm tr}\, g_{(1)})^2\bigg{]}\,.
\nonumber
\end{eqnarray}
where, \eg $R^{(0)}_{\,a b}$ corresponds to the Ricci tensor calculated
for the boundary metric $g_{(0)a b}$. Further $\epsilon$ defines a UV
regulator surface which cuts off the radial integral. The constant
coefficients appearing in the above expression \reef{holact} are given
by
 \beqa
t_1&=&\frac{1}{2}(\lambda\fin-3\mu\fin^2)=-4t_2=t_3\,,
 \nonumber\\
A&=&-(1-2 \lambda \fin-3 \mu \fin^2)=-2\, B=C\,,
\labell{coeff4}\\
D &=& -\frac{1}{6}(7-\lambda \fin   +5 \mu \fin^2 )\,,\quad
 {\cal N}=\frac{L^3}{\lp^3\fin^{3/2}}\,. \nonumber
 \eeqa
Next we eliminate $g_{(1)ij}$ using its equation of motion and then
following \cite{renorm1}, we can interpret the result as
 \be
I\simeq-\log\epsilon\,\frac{1}{2} \int d^4 x \sqrt{g_{(0)}}\ \langle
\,T_{a}{}^{a}\,\rangle\,.
 \ee
Hence comparing the coefficients of the various terms involving the
background curvatures with eq.~\reef{tranomaly}, we find
 \beqa
c&=& \pi^2\frac{L^3}{\lp^3}\frac{1}{\fin^{3/2}}\left(1-2\la\fin -3\mu\fin^2 \right)\ ,
 \labell{cc}\\
a&=&\pi^2\frac{L^3}{\lp^3}\frac{1}{\fin^{3/2}}\left(1-6\la\fin +9\mu\fin^2 \right)\ .
 \labell{aa}
 \eeqa
Given these results in eqs.~\reef{cc} and \reef{aa}, we also have
 \be
\frac{c-a}{c}=\frac{4\fin(\la-3\mu\fin)}{1-2\la\fin-3\mu\fin^2}\ .
 \labell{test1}
 \ee

\subsection{Two-point function} \label{2ptSec}

Now we turn to computing the two-point function of the stress tensor as
an alternative approach to determining the central charge $c$. It is
known \cite{Dobrev,osborn1,osborn2} that in a four-dimensional
CFT\footnote{We assume a Minkowski signature for the metric. Hence in
eq.~\reef{define8}, $x_a=\eta_{ab}x^b$ (\ie $x_0=-t$).}
 \be
\langle\, T_{a b}(x)\, T_{cd}(x')\, \rangle= \frac{C_T}{(x-x')^8}
\,{\cal I}_{ab,cd}(x-x')
 \labell{2pt}
 \ee
where
 \beqa
{\cal I}_{ab,cd}(x)&=& \frac{1}{2}\left(I_{ac}(x)
I_{bd}(x)+I_{ad}(x)I_{bc}(x)\right)-\frac{1}{4}\eta_{ab}\eta_{cd}
\nonumber\\
{\rm and}\quad&&
 I_{ab}(x)= \eta_{ab}-2 \frac{x_a x_b}{x^2}\,.
 \labell{define8}
 \eeqa
This structure is completely dictated by the constraints imposed by
conformal symmetry \cite{osborn1,osborn2}. The coefficient $C_T$ is
related to the central charge $c$ which appears as the coefficient of
the (Weyl)$^2$ term in the trace anomaly \reef{tranomaly}:
 \be
C_T=\frac{40}{\pi^4}\, c\,.\labell{relate}
 \ee

In order to compute $C_T$, it is sufficient to focus on the specific
case $\langle\, T_{xy}\, T_{xy}\, \rangle$. To determine this two-point
function, we will turn on a perturbation $r^2 h_{xy}(r,z)/L^2$ in the
AdS$_5$ background, \ie in eq.~\reef{metric0} after setting $r_0=0$.
The quadratic action for $h_{xy}=\phi$ can be written as
 \be
I_{2}=\frac{1}{2\lp^3}\int d^5 x \left( K_r (\partial_r \phi)^2 +K_z
(\partial_z \phi)^2 +\partial_r \Gamma\right)\,,\labell{acth}
 \ee
where
\begin{eqnarray}
K_r &=& -\frac{r^5 \fin^{1/2}}{2  L^5}\, (1-2\la \fin -3\mu \fin^2 )\,,
\labell{kr}\\
K_z &=& -\frac{r}{2 \fin^{1/2} L}\, (1-2\la \fin -3\mu \fin^2 )\,.
\labell{kz}
\end{eqnarray}
The details of $\Gamma$ are unimportant since the $\partial_r\Gamma$
contribution is canceled by a generalized Gibbons-Hawking boundary term
\cite{BuchelDependence}. Upon making the ansatz
 \be
\phi= e^{i p z} H_p(r)\,,
 \ee
the equation of motion for $\phi$ reduces to
 \be
H_p''(r)+\frac5r H_p'(r)-\frac{L^4 p^2}{\fin r^4} H_p(r)=0\,.
 \ee
The general solution can be written as
 \be
H_p(r)= C_1\,\frac{1}{r^2}\, K_2\! \left(\frac{L^2
p}{\fin^{1/2}r}\right) +C_2\,\frac{1}{r^2}\, I_2\!\left(\frac{L^2
p}{\fin^{1/2}r}\right) \,,
 \ee
where $I_2(x)$ and $K_2(x)$ are modified Bessel functions of the first
and second kind, respectively. In order to fix $H_p(r=\infty)=1$, we
set $C_1=p^2 L^4/(2\fin)$ and $C_2=0$. Using the equations of motion,
eq.~(\ref{acth}) can then be rewritten as
 \be
I_{2}=\frac{1}{2\lp^3} \int d^5 x\, \partial_r (K_r \phi\,
\partial_r\phi)\,.
 \ee
Substituting in the solution, we make a Fourier transform and extract
the term proportional to $\log |p|$, which amounts to ignoring all
contact terms:
 \be\labell{res1}
\langle\, T_{xy}\,T_{xy}\,\rangle(p) =\frac{L^3}{8 \lp^3 \fin^{3/2}}
(1-2\la \fin -3\mu\fin^2 )\, p^4 \log |p|\,.
 \ee

In order to compare the above result with eq.~(\ref{2pt}), it
convenient to recast the latter in Fourier space following
\cite{gubkleb}. We re-express the two-point function as \cite{osborn2}
 \be
\langle\, T_{ab}(x)\, T_{cd}(x')\, \rangle=\frac{C_T}{40}\,
\mathcal{E}^C_{\,ab}{}^{ef}{}_{,\,cd}{}^{gh}\,
\partial_e \partial_f\,\partial_g'
\partial_h'\, \frac{1}{(x-x')^4}\,,
 \labell{2ptf}
 \ee
where the tensor $\mathcal{E}^C$ satisfies
\begin{eqnarray}
\mathcal{E}^C_{ab ef \,,\,cd gh }\,k^e k^f k^g k^h&=&
 \frac{1}{24}\bigg{(}2\, k_a k_b k_c k_d-\frac{3}{2} k^2 (k_a k_c \eta_{bd}
 +k_b k_c \eta_{ad}+k_a k_d \eta_{bc}+k_b k_d \eta_{ac})
 \labell{bigmess} \\
  &&\ \ +k^2(k_a k_b \eta_{cd} +k_{c}k_{d} \eta_{ab})+
  \frac{3}{2}(k^2)^2 (\eta_{ac}\eta_{bd}+\eta_{ad}\eta_{bc})
  -(k^2)^2 \eta_{ab}\eta_{cd}\bigg{)}\,.\nonumber
\end{eqnarray}
In the case of interest with $(ab)=(xy)=(cd)$, the factor involving
$\mathcal{E}^C$ and the four derivatives simply evaluates to
 \be
p^4/16\,.
 \ee
Hence in momentum space, the two-point function can be written as
\cite{gubkleb}
 \be
\langle\, T_{xy}\,T_{xy}\,\rangle(p) =\frac{C_T}{640}\, p^4 \int d^4
x\, \frac{e^{i p\cdot x}}{x^4}=\frac{\pi^2\, C_T}{320}\, p^4\, \log
|p|+({\rm analytic~in~}p)\,. \labell{house9}
 \ee
Comparing eqs.~(\ref{res1}) and \reef{house9}, we find
 \be
C_T=\frac{40}{\pi^2}\,\frac{L^3}{\lp^3}\frac{1}{\fin^{3/2}} (1-2\la
\fin -3\mu\fin^2 )\labell{resultCT}
 \ee
and finally using eq.~\reef{relate}, we have
 \be
c= \pi^2\frac{L^3}{\lp^3}\frac{1}{\fin^{3/2}}\left(1-2\la\fin
-3\mu\fin^2 \right)\,.\labell{cc9}
 \ee
This expression precisely matches that in eq.~\reef{cc} which was found
using the holographic trace anomaly in the previous section.

\section{Holographic computation of energy fluxes} \label{flux}

At this point, our holographic dictionary contains two entries. That
is, eqs.~\reef{cc} and \reef{aa} relating the central charges of the
four-dimensional CFT to the couplings of the five-dimensional bulk
gravity theory. However, the quasi-topological gravity \reef{action} is
characterized by three independent dimensionless parameters: $\lambda$,
$\mu$ and $L/\lp$. Hence we need to extend the dictionary further by
identifying additional parameters which play an analogous universal
role in the dual CFT. Further, since we are only dealing with the
gravitational sector of the AdS$_5$ theory, we must find parameters
governing the behaviour of the stress tensor in the CFT. A natural next
step is to consider the three-point function of $T_{ij}$, as was
extensively studied by \cite{osborn1,osborn2}. There it was shown that
conformal symmetry and energy conservation are powerful enough to
determine the the three-point function up to three constants, which are
labeled $\A$, $\B$ and $\C$ in \cite{osborn2}. In fact, the two central
charges can be expressed in terms of these three parameters, as we will
elucidate below --- see eqs.~\reef{cc1} and \reef{aa1}. Further as
discussed in \cite{hofmal}, constructing a holographic model which can
explore the full range of these CFT parameters requires the
introduction of curvature-squared and curvature-cubed interactions in
the bulk gravity theory. In fact, this was the primary motivation for
constructing the quasi-topological gravity theory.

One can perform a holographic calculation of the full three-point
function \cite{rush}, however, extending these calculations to
quasi-topological gravity proves to be extremely challenging.
Therefore, we choose an indirect route to determining these
coefficients here. In particular, we construct a holographic
description of a particular thought experiment proposed for
four-dimensional CFT's in \cite{hofmal}. The experiment consists of
first producing a disturbance, which is localized and injects a fixed
energy, with an insertion of the stress tensor
$\varepsilon_{ij}\,T^{ij}$, where $\varepsilon_{ij}$ is a constant
polarization tensor. Then one measures the energy flux escaping to null
infinity in the direction indicated by the unit vector $\nvec$. The
final result takes the form
 \beq
\langle \E(\nvec)\rangle = \frac{E}{4\pi}\left[1 + t_2\,
\left(\frac{\veps^*_{ij}\veps_{ik} n^j n^k}{\veps^*_{ij}\veps_{ij}} -
\frac{1}{3}\right) + t_4\, \left(\frac{|\veps_{ij}n^i
n^j|^2}{\veps^*_{ij}\veps_{ij}}-\frac{2}{15}\right)\right]\,,
 \labell{basic}
 \eeq
where $E$ is the total energy. The structure of this expression is
completely dictated by the symmetry of the construction.  Hence the two
constant coefficients, $t_2$ and $t_4$, are parameters that
characterize the underlying CFT. Our holographic computation of $t_2$
and $t_4$ will extend our AdS/CFT dictionary to the point where the
three independent gravitational couplings will be related to three
independent parameters in the dual CFT.

Note that the (negative) constants appearing in eq.~\reef{basic} in the
two factors multiplied by $t_2$ and $t_4$ were chosen so that  these
factors contribute zero net flux when integrated over all directions.
The negative sign of these constants leads to interesting constraints
on the coefficients $t_2$ and $t_4$, which we discuss below in section
\ref{constraints}.

\subsection{Field theory calculations}

Let us first consider the discussion of \cite{hofmal} in more detail.
Again, we begin by making a small localized perturbation of the CFT in
Minkowski space with metric $ds^2=-dt^2+\delta_{ij}dx^idx^j$. With time
this perturbation spreads out, and sufficiently far away one may
imagine taking successively larger concentric two-spheres through which
one is measuring the energy flux. Let us parameterize the points on
these two spheres by a radius $r$ (\ie $r^2=x_ix^i$ as usual) and a
unit vector $\mbf n=(n^1,n^2,n^3)$. Then the energy flux measured in
the direction given by $\mbf n$ is given by
 \be
\E(\nvec)=\lim_{r\to +\infty} r^2 \int_{-\infty}^{+\infty} dt~
T^{t}{}_{i}(t,r\, \mbf n )\ n^i\,.\labell{flux0a}
 \ee
As it stands, the flux in eq.~\reef{flux0a} includes contributions from
both the past and future null boundaries of Minkowski space. To
separate out the future contribution which we are interested in, we
select one of the coordinates $x^3$ and construct light-cone
coordinates $x^\pm=t\pm x^3$. Then it is clear that the integral above
has two contributions, namely one from future null infinity
$x^+\rightarrow+\infty$ and another one from past null infinity
$x^-\rightarrow-\infty$. We will take only the former.\footnote{Note
that the following discussion overlooks the flux contribution on the
`hemisphere' at $x^-\rightarrow+\infty$. However, our primary concern
is the functional dependence of $\E(\nvec)$ and so this does not affect
our results.} For large $r$, it is convenient to write
 \be
r^2=(x^+-x^-)^2+(x^1)^2+(x^2)^2 \quad\stackrel{x^+\to
+\infty}{\longrightarrow}\quad (x^+)^2(1+ (y^1)^2+(y^2)^2), \quad
\mbox{where} \quad y^{1,2}\equiv x^{1,2}/x^+. \nonumber
 \ee
Therefore we obtain
 \be
\E(\mbf n)=-\lim_{x^+\to +\infty}(x^+)^2\left(1+(y^1)^2+(y^2)^2 \right)
\int_{-\infty}^{+\infty}dx^-\, \left[\, T_{+i}(x^+,x^-,\mbf
n)+T_{-i}(x^+,x^-,\mbf n)\,\right]\, n^i\,. \labell{flux1}
 \ee
Motivated by the preceding, we define new coordinates $y^a$:
 \be
y^+=-\frac 1{x^+}\,, \quad y^-=x^--\frac{(x^1)^2+(x^2)^2}{x^+} \,,\quad
y^{1,2}=\frac{x^{1,2}}{x^+}\,.\labell{newcoord}
 \ee
In terms of these coordinates, the desired energy flux is measured
$y^+=0$. Further it is not difficult to show that on this surface, we
have
 \be
y^{1,2}=\frac{n^{1,2}}{1+n^3}\,.\labell{bard}
 \ee
Now we transform the energy momentum tensor from $x^a$ to $y^a$
coordinates, as usual
 \be
T^x_{\,ab} = \frac{\partial y^c}{\partial x^a}\,\frac{\partial
y^d}{\partial x^b}\,T^y_{\,cd} \,.\labell{usual}
 \ee
Here we are simplifying our notation with the superscripts, $x$ and
$y$, to indicate in which coordinate system the stress tensor is
written, \ie $T^y_{\,--}=T_{y^-y^-}$. Now at $y^+=0$, we obtain
 \bea
T^x_{\, ++}&=&\left((y^1)^2+(y^2)^2\right)^2\, T^y_{\, --}\,, \quad
T^x_{\, +\,1,2}=-2\,y^{1,2}\, \left((y^1)^2+(y^2)^2\right )\,T^y_{\,
--}\,,
  \labell{usual0}\\
T^x_{\, +-}&=& \left((y^1)^2+(y^2)^2\right )\,T^y_{\, --}\,,\quad
T^x_{\, --}=T^y_{\, --}\,, \quad T^x_{\, -\,1,2}=-2\, y^{1,2}\, T^y_{\,
--}\,. \nonumber
 \eea
Hence we see that on this surface (\ie $y^+=0$), there is a single
relevant component of the energy momentum tensor in $y$ coordinates.
Using these relations, we rewrite eq.~\reef{flux1} as:
 \be
\mathcal E(\mbf n)=\Omega^{3} \,\int_{-\infty}^{+\infty} dy^-
\,\frac{T^y_{\ --}(y^+=0,y^-,y^1,y^2)}{(y^+)^2} \labell{flux5}
 \ee
with $\Omega=2/(1+n^3)$.

Further, we note that transforming from the $x^a$ to $y^a$ coordinates
produces a Weyl scaling of the metric. Hence it is natural to use this
transformation to perform the conformal transformation:
 \bea
\mathrm{d}s^2&=&-\mathrm{d}x^+ \mathrm{d}x^-+(\mathrm{d}x^1)^2+(\mathrm{d}x^2)^2=\frac{-\mathrm{d}y^+
\mathrm{d}y^-+(\mathrm{d}y^1)^2+(\mathrm{d}y^2)^2}{(y^+)^2}\nonumber\\
&&\quad \longrightarrow \qquad \mathrm{d}\tilde{s}^2=-\mathrm{d}y^+
\mathrm{d}y^-+(\mathrm{d}y^1)^2+(\mathrm{d}y^2)^2 \,. \labell{conformal0}
 \eea
Then the energy momentum tensor transforms
 \be
\tilde T_{ab}= \left|\frac{\partial x}{\partial y}\right|^{-1/2} \
\frac{\partial x^c}{\partial y^a} \, \frac{\partial x^d}{\partial
y^b}\,T_{cd} \,. \labell{conformal1}
 \ee
In particular, we have $\tilde T^y_{\,--}=T^y_{\,--}/(y^+)^2$ and
therefore eq.~\reef{flux5} becomes:
 \be
\mathcal E(\mbf n)=\Omega^{3}\int_{-\infty}^{+\infty} dy^- \ \tilde
T^y_{\, --}(y^+=0,y^-,y^1,y^2). \labell{operatorE}
 \ee

Following \cite{hofmal}, we wish to consider the expectation value of
this flux operator $\E(\nvec)$ for a particular state
 \be
\langle \mathcal E(\mbf n)\rangle= \frac{ \langle 0|\, \mathcal
O_\mt{E}^\dagger\, \mathcal E(\mbf n)\, \mathcal
O_\mt{E}\,|0\rangle}{\langle 0|\, \mathcal O_\mt{E}^\dagger\,\mathcal
O_\mt{E}\,|0 \rangle}\,.\labell{expect}
 \ee
In the present discussion, the operator $\mathcal O_\mt{E}$ is assumed
to be a localized insertion of the stress tensor of the form
 \be
\mathcal O_\mt{E}=\int d^4x\,
\varepsilon^{ij}\,T_{ij}\,e^{-iEt}\,\psi(x/\sigma)\,.
 \labell{operator}
 \ee
Here $\psi(x/\sigma)$ is some profile which localizes the insertion to
$x^a=0$ on the scale $\sigma$. We assume $E\gg1/\sigma$ and so the
energy of the insertion is $E$ up to order $1/\sigma$ corrections.
Finally since the stress tensor enters this construction, the operator
also contains a constant polarization tensor $\varepsilon_{ij}$ which
we assume only has spatial components. The symmetry of this
construction then dictates that the flux expectation value
\reef{expect} takes the form given in eq.~\reef{basic}. Further, it is
clear that by construction the result is completely determined by the
three-point function of the stress tensor. Hence the flux parameters
$t_2$ and $t_4$ appearing in eq.~\reef{basic} will be related to $\A$,
$\B$ and $\C$, the parameters controlling this three-point function.

\subsection{Holographic calculations} \label{holocalc}

The $x^a$ and $y^a$ coordinates defined in the previous section are
easily extended into the AdS$_5$ bulk with
 \bea
\mathrm{d}s^2&=&\frac{\tilde L^2}{z^2}\left(-\mathrm{d}x^+
\mathrm{d}x^-+(\mathrm{d}x^1)^2+(\mathrm{d}x^2)^2+\mathrm{d}z^2\right)\labell{bulkcoordx}\\
&=&\frac{\tilde L^2}{u^2}\left(-\mathrm{d}y^+
\mathrm{d}y^-+(\mathrm{d}y^1)^2+(\mathrm{d}y^2)^2+\mathrm{d}u^2\right)\,. \labell{bulkcoordy}
 \eea
Recall that $\tilde L$ is the curvature of the AdS$_5$ geometry, as
defined in eq.~\reef{curf}. To relate these two coordinate systems, it
is convenient to describe AdS$_5$ as the hyperbola
 \be
-(W^{-1})^2-(W^0)^2+(W^1)^2+(W^3)^2+(W^3)^2+(W^4)^2=-\tilde{L}^2
\labell{Wparam}
 \ee
in a six-dimensional Minkowski space with (--,--,+,+,+,+) signature.
Note that we reach the boundary of AdS$_5$ by taking $W^M$ large. The
previous coordinates are mapped to the $W^M$ coordinates with
 \bea
 W^{-1}+W^4 &=& \frac{\tL^2}{z}\,, \qquad W^a=\tL\, \frac{x^a}{z}\
 \quad
{\rm for}\ a=0,1,2,3\,, \labell{www}\\
W^0+W^3&=&\frac{\tL}{u}\,, \quad W^{-1}+ W^{4}=
-\tL^2\,\frac{y^+}{u}\,,\quad W^{-1}- W^{4}= - \frac{y^-}{u}\,, \quad
W^{1,2}= \tL\,\frac{y^{1,2}}{u}\,.\nonumber
 \eea
Note that $z$ and $u$ are mapped to two orthogonal null surfaces in the
$W^M$ space. Further the powers of $\tL$ are slightly different in the
second line above to ensure that the (engineering) dimension of the
coordinates is properly accounted for, \ie $u$ is dimensionless while
$y^+$ has dimensions {\it length}$^{-1}$. With eq.\reef{www}, we can
relate the $(x^a,z)$ and $(y^a,u)$ coordinate systems in
eqs.~\reef{bulkcoordx} and \reef{bulkcoordy} as
 \bea
y^+&=& -\frac 1{x^+}\,, \qquad y^{1,2}= \frac{x^{1,2}}{x^+}\,, \qquad
u=\frac{z}{x^+}\,,\nonumber\\
y^-&=&x^--\frac{(x^1)^2+(x^2)^2}{x^+}-\frac{z^2}{x^+}\,.\labell{xzyu}
 \eea
Notice that on the asymptotic boundary $z=0$, the above coordinate
transformation reduces to that given in eq.~\reef{newcoord}. Further
with $y^+=0$ and any finite value of $u$, we are on the AdS$_5$ horizon
at $z=\infty$ in the $(x^a,z)$ coordinates.

As commented above, in calculating the flux expectation value in
eq.~\reef{expect}, we are essentially determining a specific component
of the  three-point function of the stress tensor. Hence in our
holographic description, we must first introduce appropriate metric
perturbations $h_{\mu\nu}$ in the AdS$_5$ bulk which couple to the dual
insertions of $T_{ab}$. We then evaluate the on-shell contribution to
the cubic effective action for the graviton insertions.

As discussed in \cite{quasi}, in general, the equations of motion for
quasi-topological gravity involve higher derivatives. Hence one would
expect that linearized equations of motion for the metric perturbations
here are also higher order. However, it was observed in \cite{quasi}
that in fact these linearized equations for gravitons propagating in
the AdS$_5$ vacuum match precisely the second order equations of
Einstein's theory, up to some overall (constant) coefficient, as shown
in eq.~\reef{fullEin}. This makes the following calculations much
simpler as we may borrow previous results \cite{hofmal,rush} for the
graviton solutions in Einstein gravity. Hence while the higher
derivative contributions in quasi-topological gravity are essential to
producing a nonvanishing value for $t_4$ in the dual CFT, they only
contribute through the three-point interactions in the following.

We first consider the flux operator $\mathcal E(\mbf n)$ in
eq.~\reef{operatorE}. It is natural to use the $(y^a,u)$ coordinates in
eq.~\reef{bulkcoordy} and the standard AdS/CFT dictionary advises us
that $\tilde T^y_{\ --}(y)$ couples to $h_{++}(y^a,u=0)$. Considering
first a localized insertion $h_{++}(y^a,u=0)
=\delta(y^1)\delta(y^2)\delta(y^+)$, the bulk solution is given by
 \be
h_{++}(y^+,y^-,y^1,y^2,u)=\frac{u^2}{(u^2-y^+(y^-
-y'^-)+(y^1)^2+(y^2)^2)^4}\,.\labell{localplusplus}
 \ee
As noted above, we are using the same solution here as in \cite{hofmal}
because the linearized equations of motion for perturbations around
$AdS_5$ in quasi-topological gravity are the same for Einstein gravity
\cite{quasi}. To obtain the operator $\mathcal E(\mbf n)$, we then
integrate in $y^-$, as well as multiplying by an overall factor of
$\Omega^3$ and performing a translation in $y^1$ and $y^2$, to obtain
 \be
h_{++}(y^+,y^1,y^2,u)=\frac{8\,\delta(y^+)}{(1+n_3)^3}\,
\frac{u^2}{\left(u^2+(y^1-y'^{1})^2+(y^2-y'^{2})^2\right)^3}
\,,\labell{hpp}
 \ee
with $y'^{1,2}=n^{1,2}/(1+n^3)$, as in eq.~\reef{bard}.

In fact we can make this insertion at a nonlinear level in the bulk
gravity theory, following \cite{hofmal} and \cite{hofman}. To achieve
this, we consider the shockwave background:
 \be
ds^2=\frac {\tL^2}{u^2}\Big[\,\delta(y^+)\, \W(y^1,y^2,u)(dy^+)^2-dy^+
dy^-+ (dy^1)^2+ (dy^2)^2+du^2\Big]
 \labell{shocking}
 \ee
This metric solves the full equations of motion coming from
eq.~\reef{action} provided that $\W(y^1,y^2,u)$ satisfies the equation
of motion
 \be
\partial_{u}^{2\,} \W-\frac{3}{u}\,\partial_u  \W+\partial_{y^1}^2
\W+\partial_{y^2}^2\W=0\,.
 \ee
This simple linear equation appears as the equation of motion in
Einstein gravity and one can readily show that it is not corrected by
the higher curvature terms in eq.~\reef{action} with the arguments of
\cite{gary}. From our expression for $h_{++}$ in eq.~\reef{hpp}, the
relevant wavefunction is
 \be
 \W(y^1,y^2,u)=\frac{\Omega^3}{\tL^2}\,
\frac{u^4}{\left(u^2+(y^1-y'^{1})^2+(y^2-y'^{2})^2\right)^3}
 \labell{solW}
 \ee
with $y'^{1,2}=n^{1,2}/(1+n^3)$, as before.

Next we turn to the graviton perturbations dual to the operator
insertion $\mathcal O_\mt{E}$ in eq.~\reef{operator}. To simplify the
discussion, we choose a particular polarization with
$\varepsilon_{x^1x^2}=1=\varepsilon_{x^2x^1}$ and all other components
vanishing. Using the $(x^a,z)$ coordinate system in
eq.~\reef{bulkcoordx}, the desired operator \reef{operator} is sourced
by a metric perturbation with the boundary value:
$h_{x^1x^2}(x^a,z\to0)=z^{-2}e^{-i E(x^++x^-)/2}$. The bulk solution
that corresponds to this boundary perturbation is then
 \be
h_{x^1x^2}(x^a,z)=\int d^4 x' \,e^{-i \frac{E}2(x'^++x'^-)}\,\frac{1
}{(z^2+(x-x')^2)^2}\,.\labell{origami}
 \ee
Since the $h_{++}$ perturbation is completely localised at $y^+=0$, for
later purposes, we will primarily be interested in the behaviour of
$h_{x^1x^2}(x^a,z)$ on that surface. Following \cite{hofmal}, it is
possible to perform the above integral using the parameterization of
AdS$_5$ in eq.~\reef{Wparam} to produce
 \be
h_{x^1x^2}(W^+\simeq0,W^-,W^i)\simeq \frac{(W^+)^2}{\tL^4E^2}\,e^{-i E
W^-/2} \delta^{3}(W^i)\,,
 \ee
where $W^\pm=W^{-1}\pm W^4$. Implicitly, $W^0$ has been replaced with
$(W^0)^2=1- (W^i)^2$ which is the reduction of eq.~\reef{Wparam} with
$W^+=0$. Using eq.~\reef{www}, we may express the coordinate dependence
in terms of $(y^a,u)$
 \be
h_{x^1x^2}(y^+\simeq0,y^-,y^1,y^2,u)\simeq \frac{(y^+)^2}{E^2}\, e^{i E
y^-/2} ~\delta(y^1)~\delta(y^2)~\delta(u-1)\,. \labell{tensorsol}
 \ee
Finally with the coordinate transformation \reef{xzyu}, we also
transform the tensor indices to find that at $y^+\to 0$, our metric
perturbation becomes
 \beq
h_{y^1y^2}\simeq \frac{1}{E^2}\, e^{i E y^-/2}
~\delta(y^1)~\delta(y^2)~\delta(u-1) \labell{hmmm}
 \eeq
along with other $h_{y^+y^1}$, $h_{y^+y^2}$ and $h_{y^+y^+}$
components. However, the form of the latter will not be important, as
we now discuss. Note that the original expression \reef{origami} was
transverse and traceless in the ($x^a,z$) coordinates but as a result,
the expression produced by simply making a coordinate transformation to
the ($y^a,u$) coordinates is not. However, it is convenient to work in
this gauge since a great simplification results in the equation of
motion for the graviton propagating in the AdS$_5$ background, \ie away
from the shockwave deformation in eq.~\reef{shocking}. Hence at this
point, we choose add to eq.~\reef{hmmm} the components required to
impose transverse traceless gauge in the ($y^a,u$) coordinates.

The mode above was traceless by construction and so we only need to
ensure that the transverse condition is satisfied as well, \ie
$\nabla^\mu h_{\mu\nu}=0$. In the present case, the latter can be
satisfied as long as $h_{y^1y^2}$ is accompanied by modes satisfying:
 \bea
\partial_{y^-} h_{y^+y^1}&=&\frac 12\,\partial_{y^2}h_{y^2 y^1}\,,
\quad
\partial_{y^-} h_{y^+y^2}=\frac 12\,\partial_{y^1}h_{y^1 y^2}\,,
\nonumber\\
\partial_{y^-}h_{y^+y^+}&=&\frac 12\left(
\partial_{y^1} h_{y^1y^+}+ \partial_{y^2} h_{y^2y^+}\right)\,.
 \labell{xmodes}
 \eea
Together the $h_{y^1y^2}$, $h_{y^+ y^1}$, $h_{y^+y^2}$ and $h_{y^+
y^+}$ components form an independent transverse traceless mode. We have
verified that with $h_{y^1y^2}(y,u)\equiv \tL^2/u^2\, \phi(y,u)$ and
the remaining components chosen to satisfy eq.~\reef{xmodes}, the
equation of motion for $\phi(y,u)$ becomes simply that of a massless
scalar in AdS$_5$ (up to interaction terms with the shockwave):
 \be
\partial_{u}^2 \phi-\frac 3u \,\partial_u \phi +
\partial_{y^1}^2 \phi+\partial_{y^2}^2 \phi-4\partial_{y^+}
\partial_{y^-}\phi=0\,.
 \ee

To find the three-point function, we add these perturbations to the
metric \reef{shocking} and evaluate the action \reef{action} on-shell.
Then we must extract the terms of the form  $\W\, \phi^2$ from this
result. After integration by parts and using the equations of motion,
the cubic effective action becomes
 \be
S^{(3)}_{\W \phi^2}=-\frac {1}{8\lp^3}\int d^5 x \sqrt{-g} ~\phi
~\partial_{-}^2\phi~ \W\Big[ 1-2 \ff \la-3 \mu \ff^2+ \ff(\la-87 \ff
\mu)\, T_2+21 \ff^2 \mu\, T_4\Big] \labell{close}
 \ee
where
 \bea
T_2 &=&  \left. \frac{\partial_1^2 \W+\partial_2^2\W-2\, \partial_u \W}{\W}
\right|_{u=1,\,y^1=y^2=0}\,,\nonumber \\
T_4 &=& \left.\left( 3\, T_2+ \frac{ \partial_1^2
\partial_2 ^2 \W-\, \partial_u \partial_1^2 \W-\,\partial_u
\partial_2^2 \W}{\W}\right)\right|_{u=1,\,y^1=y^2=0}\,.
 \labell{coeffT}
\eea
Implicitly here, we are using that with the perturbations given above,
\ie eqs.~\reef{hpp} and \reef{tensorsol}, the interaction is entirely
localized along $y^+=0=y^1=y^2$ and $u=1$. Substituting the solution
\reef{solW} for $\W(y^1,y^2,u)$ into eq.~\reef{coeffT}, we obtain
 \be
T_2=24\left (\frac{n_1^2+n_2^2}{2}-\frac 13\right)\,\ \  \qquad
T_4=180\left(2\, n_1^2\, n_2^2-\frac 2{15}\right)\,.
 \labell{closer}
 \ee
The expressions above involving $n^1$ and $n^2$ should be interpreted
as two independent $SO(3)$-invariant combinations of the unit vector
$n^i$ and the (implicit) polarization tensor $\varepsilon_{ij}$, \ie
 \be
\frac{n_1^2+n_2^2}{2} =\frac{\veps^*_{ij}\,\veps_{ik}\, n^j
n^k}{\veps^*_{ij}\,\veps_{ij}}\,,\qquad
2 n_1^2 n_2^2 = \frac{|\veps_{ij}\,n^i
n^j|^2}{\veps^*_{ij}\,\veps_{ij}}\,.
 \labell{combo1}
 \ee
This was the guiding principle in selecting the two combinations
presented in eq.~\reef{coeffT}. To normalize the final result, we must
divide by the two-point function $\langle T_{y^1y^2}\,
T_{y^1y^2}\rangle$, which is essentially the calculation of section
\ref{2ptSec}. We finally arrive at an expression identical to that in
eq.~\reef{basic} with
 \be
t_2=\frac{24\fin(\lambda-87 \fin \mu)}{1-2 \ff \lambda-3 \ff^2 \mu}\,
,\qquad t_4=\frac{3780 \ff^2 \mu}{1-2 \ff \lambda-3 \ff^2 \mu}\,.
 \labell{t2t4}
 \ee

\subsection{General three-point parameters} \label{diction3}

At this point, we return to $\A$, $\B$ and $\C$, the parameters in the
dual CFT controlling the general structure of the three-point function
of the stress tensor \cite{osborn1,osborn2}. In fact, these parameters
fix both the central charges, $c$ and $a$, and the flux parameters,
$t_2$ and $t_4$, in the CFT. Hence we can use the results in the
previous section and in section \ref{central} to express the
three-point parameters in terms of the gravitational couplings.

First, we can express the central charges as \cite{osborn1}
 \beqa
c&=& \frac{\pi^6}{480}\left(9\A-\B-10\C\right)\ ,
 \labell{cc1} \\
a&=& \frac{\pi^6}{2880}\left(13\A-2\B-40\C\right)\ .
 \labell{aa1}
 \eeqa
Further we have \cite{hofmal,GBanyd}
 \be
t_2=\frac{15(5\A+4\B-12\C)}{9\A-\B-10\C}\ ,\qquad
t_4=-\frac{15(17\A+32\B-80\C)}{4(9\A-\B-10\C)}\ . \labell{t2t4b}
 \ee
Given that these four quantities are all determined by the same three
parameters, these expressions must be redundant. That is, we can see
that there is a consistency condition:
 \be
\frac{c-a}{c} = \frac 16 t_2 + \frac 4{45} t_4=
\frac{41\A-4\B-20\C}{6(9\A-\B-10\C)}\ .
 \labell{consist}
 \ee
With the results \reef{t2t4} in the previous section,
 \be
\frac 16 t_2 + \frac 4{45} t_4
=\frac{4\fin(\la-3\mu\fin)}{1-2\la\fin-3\mu\fin^2}\,.
 \labell{test2}
 \ee
Now comparing to eq.~\reef{test1}, we see that our holographic results
satisfy the required relation \reef{consist}.

Combining these expressions (\ref{cc1}--\ref{t2t4b}) with the results
of the holographic calculations, eqs.~\reef{t2t4}, \reef{cc1} and
\reef{aa1}, we arrive at the following expressions
 \beqa
\A&=& -\frac{512}{9\pi^4}\frac{L^3}{\lp^3}\frac{1}{\fin^{3/2}}
\left(1-12\la\fin+48\mu\fin^2\right)\ ,
 \labell{A1} \\
\B&=& -\frac{32}{9\pi^4}\frac{L^3}{\lp^3}\frac{1}{\fin^{3/2}}
\left(49-318\la\fin+4377\mu\fin^2\right)\ ,
 \labell{B1} \\
\C&=& -\frac{32}{9\pi^4}\frac{L^3}{\lp^3}\frac{1}{\fin^{3/2}}
\left(23-168\la\fin+213\mu\fin^2\right)\ .
 \labell{C1}
 \eeqa

\section{Physical constraints} \label{harmsway}

Having established several interesting entries in the AdS/CFT
dictionary for quasi-topological gravity, we next consider various
constraints on the gravitational couplings that arise to ensure the
physical consistency of the dual CFT's. We consider three independent
constraints in the following:

\subsection{Positivity of $C_T$} \label{positiveCT}

Unitarity of the CFT requires that the central charge $C_T$ or $c$ is
positive. This constraint on $c$ may seem somewhat mysterious from the
point of view of the holographic trace anomaly discussed in section
\ref{trace}. However, the definition in section \ref{2ptSec} shows
$C_T$ appears in the two-point function and so this central charge
controls the sign of the norm of CFT states created with the stress
tensor. Hence given eq.~\reef{resultCT}, the dual gravity theory must
satisfy
 \be
1-2 \fin\, \lambda-3\fin^2\, \mu>0\,. \labell{CTposit}
 \ee

At this point, we note that multiplying the expression on the left-hand
side of eq.~\reef{CTposit} by minus one yields precisely the derivative
of the left-hand side of eq.~\reef{fin}, \ie the slope of the
polynomial there evaluated at the root given by eq.~\reef{fin}. This
comment is related to the observation in \cite{quasi} that this slope
appears as a pre-factor in the kinetic term for graviton or in the
linearized equations of motion \reef{fullEin} in the AdS vacua of the
theory. That is, the sign of the slope determines whether or not the
graviton is a ghost in a particular AdS vacuum --- the graviton is
well-behaved when the slope is negative. Of course, the holographic
calculation in section \ref{2ptSec} shows that $C_T$ is precisely
determined by the graviton propagator and so it is no surprise that the
same factor appears in both places. Further, we note that in AdS vacua
with a ghost-like graviton, this pathology would make a prominent
appearance as non-unitarity in the dual CFT since $C_T$ would be
negative. The analogous observations were made for GB gravity in
\cite{GBanyd}.

\subsection{Positivity of energy fluxes} \label{constraints}

Turning to the expression of the energy flux in eq.~\reef{basic}, we
note that the two factors multiplied by $t_2$ and $t_4$ were normalized
to give a vanishing contribution to the net flux when integrated over
all directions. Hence depending on the specific direction, these
factors may give either a positive or negative contribution to $\langle
\E(\nvec) \rangle$. Further, it is easy to see that if the coefficients
$t_2$ and $t_4$ become too large, the energy flux measured in various
directions will become negative. Following \cite{hofmal}, avoiding this
problem then imposes various constraints on these
coefficients:\footnote{A complementary analysis in \cite{huge3}
produced a constraint equivalent to eq.~\reef{spin0a}, again as a
positivity constraint on the three-point function of the stress
tensor.}
 \bea
{\rm Tensor:}&&1-\frac 13 t_2- \frac 2{15} t_4 \geq 0\,,  \labell{spin2a}\\
{\rm Vector:}&&1+\frac 16 t_2- \frac 2{15} t_4 \geq 0\,,
\labell{spin1a}\\
{\rm Scalar:}&&1+\frac 13 t_2+ \frac 8{15} t_4 \geq 0\,.
\labell{spin0a}
 \eea
If the polarization tensor is chosen with
$\varepsilon_{x^1x^2}=1=\varepsilon_{x^2x^1}$ and all other components
vanishing, as in section \ref{holocalc}, the tensor, vector and scalar
constraints above correspond to demanding a positive flux with
$n_3^2=1$, $n_1^2=1$ (or $n_2^2=1$), and $n_1^2=1/2= n_2^2$,
respectively. Using eq.~\reef{t2t4}, we translate these constraints on
$t_2$ and $t_4$ to constraints on the gravitational couplings:
 \bea
{\rm Tensor:}&&1-10 \ff \lambda+189 \ff^2 \mu\geq 0\,,
\labell{spin2}\\
{\rm Vector:}&&1+2 \ff \lambda-855 \ff^2 \mu\geq 0\,,
\labell{spin1}\\
{\rm Scalar:}&& 1+6 \ff \lambda+1317 \ff^2 \mu\geq 0\,. \labell{spin0}
 \eea
In the present case, these constraints confine the higher curvature
couplings of quasi-topological gravity to lie within a small region in
the ($\lambda,\mu$)-plane, as shown in figure \ref{3ptplot}.
\FIGURE{ \centering
		\includegraphics[width=120mm]{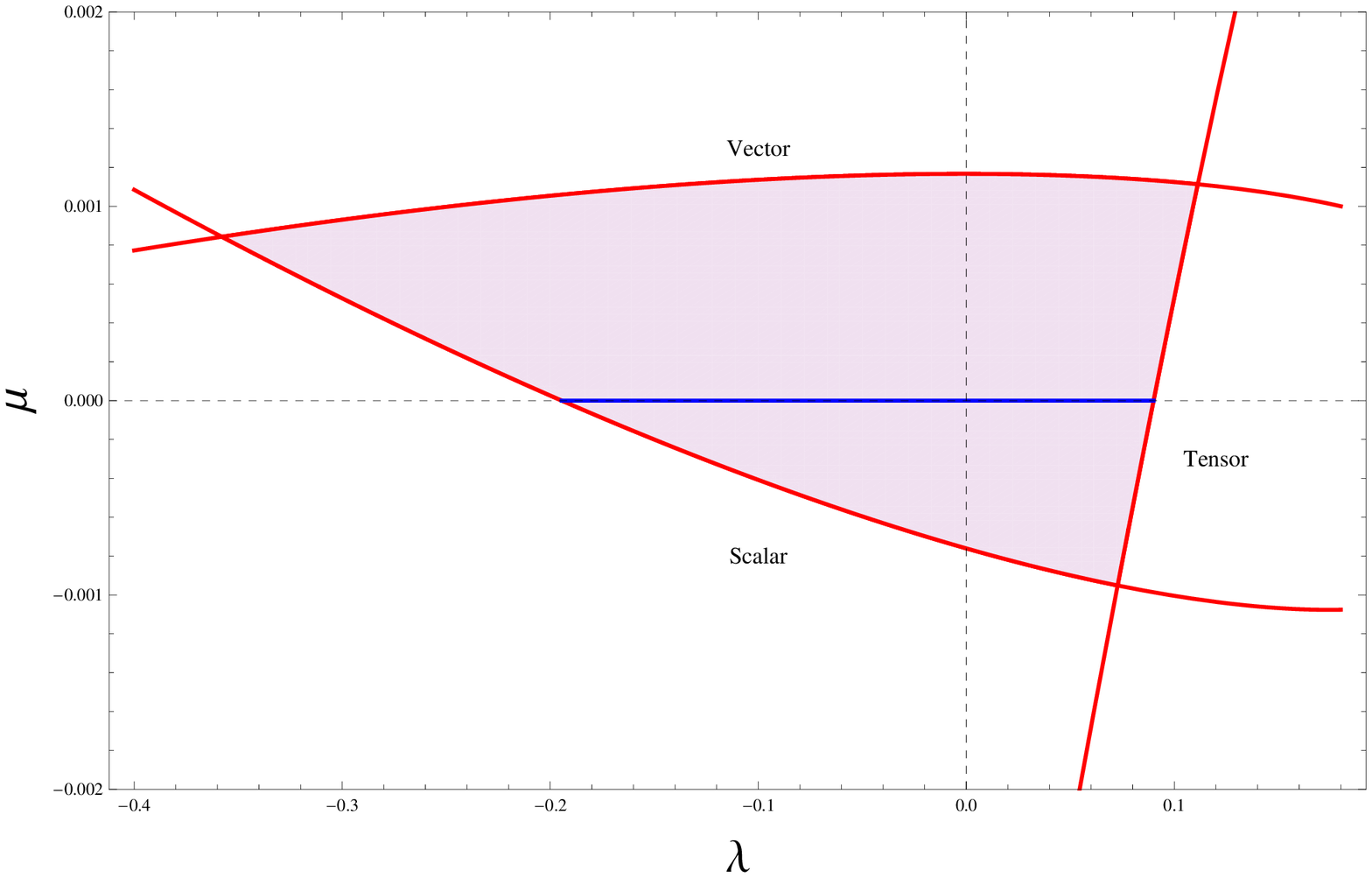} \caption{The
allowed region in ($\lambda,\mu$)-plane satisfying the constraints
appearing in eqs.~(\ref{spin2}--\ref{spin0}). Within this region, the
energy flux \reef{basic} in the dual CFT is positive for any direction.
The (blue) segment on the $\lambda$-axis within the allowed region
matches precisely the allowed values of the coupling in
five-dimensional GB gravity. } \labell{3ptplot} }

Setting $\mu=0$ reduces the theory to GB gravity and one recovers the
expected constraints from eqs.~(\ref{spin2}--\ref{spin0}) in this limit
\cite{alex0,hofman}. First with $\mu=0$, eq.~\reef{fin} yields
$\fin=\frac{1}{2\lambda} \left(1- \sqrt{1-4\lambda} \right)$ for the
ghost-free AdS vacuum. Then, for example, the tensor constraint
\reef{spin2} reduces to $5\sqrt{1-4\lambda}-4\geq 0$ or $\lambda\le
9/100$. Similarly, the vector \reef{spin1} and scalar \reef{spin0}
constraints yield $\lambda\geq-3/4$ and $\lambda\geq-7/36$,
respectively. Hence to maintain positive energy fluxes in all
directions, the curvature-squared coupling must lie in the range
$-\frac{7}{36}\le\lambda\le\frac{9}{100}$, as expected
\cite{alex0,hofman}.  While the latter combines the results for the
tensor and scalar constraints, the inequality arising from the vector
constraint also matches the previously derived result
\cite{alex0,hofman}. The allowed GB theories are illustrated in figure
\ref{3ptplot} as the blue segment on the $\lambda$-axis (\ie $\mu=0$)
within the allowed region.

\subsection{Causality constraints} \label{causalconstr}

Constraining the gravitational couplings by demanding that the dual CFT
respects causality was first explored in the context of
five-dimensional Gauss-Bonnet gravity \cite{shenker1,shenker2}. In this
analysis, one considers graviton fluctuations that probe the bulk
geometry. The AdS$_5$ vacuum (\ie eq.~\reef{metric0} with $f(r)=\fin$)
is Lorentz invariant in the CFT directions and so no violations of
causality would be found with this bulk spacetime. Instead, the black
hole solution provides a background where Lorentz invariance is broken
and in certain instances, the dual CFT plasma supports superluminal
signals. Hence one constrains the gravitational couplings to avoid the
appearance of such superluminal modes. The original analysis
\cite{shenker1,shenker2} of GB gravity only considered gravitons
polarized transversely to the momentum direction, in what is
conventionally called the tensor channel. The analysis was later
extended to the shear and sound channels in \cite{alex0,hofman}. These
causality constraints have since been extended to GB gravity in higher
dimensions \cite{GBanyd,spain1,GBrefs} and more generally to higher
order Lovelock theories \cite{LLrefs}.

In all of these cases, it was found that the causality constraints
precisely match those arising from requiring positive energy fluxes. In
particular, for five-dimensional GB theory, these constraints are
exactly equivalent to those presented in
eqs.~(\ref{spin2}--\ref{spin0}) with $\mu=0$. However, it has been
shown that this matching does not appear in general, in particular for
cases where the gravitational equations of motion are not second order
\cite{hofman}. Hence, in general, one has two independent sets of
constraints, one required by the positive fluxes and a second
determined by the absence of superluminal modes. In quasi-topological
gravity, the linearized equations in a general background (and as we
will see, in a black hole background) are fourth order in derivatives
and so we do not expect that the previous constraints
(\ref{spin2}--\ref{spin0}) will be reproduced by the causality
analysis. However, unfortunately our final results here will be similar
to those in \cite{hofman}. That is, we find no evidence of causality
violation once the curvature-cubed coupling $\mu$ is turned on in
quasi-topological gravity.

There is a broad literature discussing causality in general field
theories \cite{fox}. The key property characterizing how quickly
signals propagate is the speed with which a wave-front propagates out
from a discontinuity in some initial data. This front velocity is given
by
\begin{equation}
v^{front}\equiv\lim_{|q|\to\infty}\ {\rm Re}(\omega)/{q} \,. \labell{cs}
\end{equation}
That is, we are interested in the phase velocity of modes in the limit
of infinitesimally short wavelengths. Hence in a relativistic field
theory, we would require that $v^{front}\le1$ in order to avoid any
acausal behaviour.

In the present holographic framework, we need to determine the front
velocity of signals in the dual CFT. That is, we must determine this
velocity for excitations dual to various graviton channels in the bulk
spacetime. Consider the black hole background given in
eq.~\reef{metric0} and define the new coordinate $\rho=r_0^2/r^2$. The
metric becomes
\beq
\mathrm{d}s^2 = \frac{r_0^2}{ L^2\rho}\left(- \frac{f(\rho)}{\fin}
\,\mathrm{d}t^2 +\mathrm{d}x_1^2+\mathrm{d}x_2^2+\mathrm{d}x_3^2\right)
+\frac{L^2}{4\rho^2 f(\rho)}\, \mathrm{d}\rho^2\,. \labell{bgr}
 \eeq
For simplicity, we will focus on tensor perturbations of the
form
 \be
h_{x^1 x^2}=\frac{r_0^2}{L^2 \rho}e^{-i \omega t+i q x_3}\phi(\rho) \labell{bang}
 \ee
propagating on the background given in eq.~\reef{bgr}. The
determination of the front velocity for such modes was described in
detail in \cite{shenker2}. For our present purposes, it suffices to
derive the effective speed in the CFT directions at large momentum and
frequency by focusing on the contributions coming from the $t$ and $z$
derivatives in the linearized equations of motion. That is, the full
linearized equation for $\phi(\rho)$ takes the form
 \be
\partial_\rho\left(\mathcal C^{(2)}(\rho,q^2)\,\partial_\rho \phi(\rho)\right)
+\mathcal C^{(0)}(\rho,q^2,\omega^2)\, \phi(\rho)=0\,,
 \labell{genZ0}
 \ee
The function $\mathcal C^{(2)}$ contains two terms, one independent of
$q$ and the other proportional to $q^2$. In contrast, the $\mathcal
C^{(0)}$ function is a sum of terms proportional to $\omega^2, q^2,
\omega^2 q^2$ and $q^4$. In the large momentum and frequency limit, the
radial derivatives can be neglected and essentially only the $\mathcal
C^{(0)}$ term is relevant above. By setting this term to zero, one
finds an effective `dispersion relation' relating the frequency to the
momentum. As the final result is quite a complicated expression, let us
approach it in several steps.

We start by setting both $\lambda$ and $\mu$ to zero, in which case our
theory reduces to Einstein gravity.  Then the dispersion relation is
simply
 \be
0=\omega^2-\frac{f(\rho)}{\fin}\,q^2\,. \labell{dispE}
 \ee
For Einstein gravity, $f(\rho)=1-\rho^2$ (and $\fin=1$). Therefore the
pre-factor multiplying $q^2$ above is less than one for any finite radius
and we expect that these excitations always propagate at less than the speed of light,
\ie $\omega^2/q^2\le1$. Next we consider GB gravity with $\lambda\ne0$
and $\mu=0$. In this case the dispersion relation becomes
\be 0=\omega^2\left(1-2\lambda f(\rho)+2\rho \lambda
f'(\rho)\right)-\frac{f(\rho)}{\fin}\,q^2 \left(1-2\lambda f(\rho)+2 \lambda \rho
f'(\rho)-4\rho^2 \lambda f''(\rho)\right)
 \ee
Using the GB black hole solutions and expanding the above expression
near the AdS boundary yields
 \be
\frac{\omega^2}{q^2}=1-\frac{1-10 \ff \lambda}{\ff (1-2 \ff
\lambda)^2}\, \rho^2+ \mathcal O\!\left(\rho^4\right)\,.
\labell{dispGB}
 \ee
Hence we expect that preventing a superluminal front velocity requires $1-10\ff
\lambda \geq 0$, which matches precisely the tensor constraint
\reef{spin2} with $\mu=0$. Of course, this agreement for GB gravity was
previously noted \cite{hofmal,alex0,hofman}.

Now for the full theory with both $\lambda\ne0$ and $\mu\neq 0$, the
story is quite different. As described in \cite{quasi}, the linearized
equations of motion describing gravitons in a general background for
quasi-topological gravity are fourth order in derivatives. As a result,
one finds that the equations for the tensor perturbations in the black
hole background now involve higher powers of momentum, \ie terms
proportional to $\omega^2 q^2$ and $q^4$, as well as $q^2
\partial_u^2$. These quartic momentum terms will then
dominate the large $q$ limit. Indeed, in this case the full effective
dispersion relation becomes
 \bea
0&=&\omega^2\left(1-2\lambda f(\rho)+2\rho \lambda
f'(\rho)\right)-\frac{f(\rho)}{\fin}\,q^2 \left(1-2\lambda f(\rho)+2 \lambda \rho
f'(\rho)-4\rho^2 \lambda f''(\rho)\right)\nonumber\\
&&\ \ -3\mu \, \omega^2\, \left[f(\rho) \left(f(\rho)-2\rho f'(\rho)+27
\rho^2 f''(\rho)+48 \rho^3 f^{(3)}(\rho)+12\rho^4 f^{(4)}(\rho)\right)
\right.\nonumber\\
&&\ \ \qquad\qquad\qquad\quad\left.+3\rho^2 f'(\rho) \left(
f'(\rho)+6 \rho \, f''(\rho)+2 \rho^2 f^{(3)}(\rho)\right)+6 \rho^4 f''(\rho)^2\right]
\nonumber\\
&&\ \ +3\mu \,  \frac{f(\rho)}{\fin}\,q^2\, \left[f(\rho) \left(f(\rho)-2\rho
f'(\rho)-23 \rho^2 f''(\rho)-48 \rho^3 f^{(3)}(\rho)-12\rho^4 f^{(4)}(\rho)\right)
\right.\nonumber\\
&&\ \ \qquad\qquad\qquad\quad\left.+\rho^2 f'(\rho) \left(
f'(\rho)-24 \,\rho\, f''(\rho)-12 \rho^2 f^{(3)}(\rho)\right)\right]
\nonumber\\
&&\ \ -12\mu\, \rho^2\frac{f(\rho)}{\fin}\, q^2 \bigg( f'(\rho)+2 \rho
f''(\rho)\bigg) \bigg( \omega^2-\frac{f(\rho)}{\fin}q^2\bigg)
\labell{fulldisp}
 \eea
Hence as commented above, in the limit of large $q$, the contributions
from the higher momentum terms, appearing in the last line above, come
to dominate. In fact then, the dispersion relation reduces to that
found for Einstein gravity \reef{dispE}. Therefore we conclude that the
dual excitations always propagate at less than the speed of light. A
similar result was also found in \cite{hofman} when considering a
Weyl-tensor squared interaction added to the usual Einstein action.

However, one might find this analysis somewhat suspect. In particular,
this effective dispersion relation may not be well-defined here because
of the very presence of these higher derivative terms which produced
the simplification in the final step. Further in these higher
derivative contributions, the asymptotic behaviour of the factor
$f''(r)+\frac{2}{r} f'(r)\sim (r_0/r)^6$ gives a very rapid decay and
so perhaps modes propagating very close to the AdS boundary can evade
our previous conclusion.

To eliminate these potential loopholes, we now proceed with a more
careful analysis by rewriting the equation of motion \reef{genZ0} in a
Schr\"odinger form, following \cite{shenker2,alex0}. The first step is
to isolate the $\omega^2$ contributions by rewriting eq.~\reef{genZ0}
as
 \be
A(\rho,\scrq^2)\, \partial_\rho^2\phi(\rho)+B(\rho,\scrq^2)\,
\partial_\rho \phi(\rho)+C(\rho,\scrq^2)\, \phi(\rho)+
D(\rho,\scrq^2)\,\scrw^2 \phi(\rho)=0
\,. \labell{PreSchro}
 \ee
Here we have defined the dimensionless frequency and momentum,
 \be
\scrw=\frac{\omega}{2\pi T}\,, \qquad \scrq=\frac{q}{2\pi T}\,.
\labell{useful1}
 \ee
Performing a change of coordinates and rescaling $\phi(\rho)= Z(\rho)
\psi(\rho)$ according to
 \bea
\frac{dy}{d\rho}= \sqrt{\frac{D(\rho,\scrq^2)}{A(\rho,\scrq^2)}}\,,
\qquad \frac{\partial_\rho Z(\rho,\scrq^2)}{Z(
\rho,\scrq^2)}=\frac{\partial_\rho A(\rho,\scrq^2)}{4\,
A(\rho,\scrq^2)}-\frac{\partial_\rho D(\rho,\scrq^2)}{4\,
D(\rho,\scrq^2)}-\frac{B(\rho,\scrq^2)}{2\, A(\rho,\scrq^2)}
\,,\labell{useful2}
 \eea
the eq.~\reef{PreSchro} takes the desired form
 \be
-\frac{1}{\scrq^2}\, \partial_y^2 \psi(y)+U(y,\scrq^2)\,
\psi(y)=\alpha^2\,\psi(y)\,,\labell{schro99}
 \ee
where $\alpha^2=\scrw^2/\scrq^2$. Note that in terms of the
Schr\"odinger coordinate $y$, the horizon now appears at $y\rightarrow
+\infty$ and the asymptotic AdS boundary, at $y=0$. In terms of the
radial coordinate $\rho$, the effective potential can be expressed as
 \be
\scrq^2 U(\rho,\scrq^2)=-\frac{A(\rho,\scrq^2)}{D(\rho,\scrq^2)}\,\frac{\partial^2_\rho
Z(\rho,\scrq^2)}{Z(\rho,\scrq^2)}-\frac{B(\rho,\scrq^2)}{D(\rho,\scrq^2)}\,\frac{\partial_\rho
Z(\rho,\scrq^2)}{Z(\rho,\scrq^2)}-\frac{C(\rho,\scrq^2)}{D(\rho,\scrq^2)}\,.
 \labell{potu1}
 \ee
More concretely the effective potential is of the form,%
 \be
\scrq^2 U(\rho,\scrq^2)=\frac{\sum_{i=0}^{i=6} n_i(u)
(\scrq^2)^i}{\sum_{i=0}^{i=5} d_i(u) (\scrq^2)^i} \labell{U exp}
 \ee
for some complicated functions $n_i$ and $d_i$. If we now take the
large momentum limit we obtain the result
 \be
U(\rho,\scrq^2)=\frac{f(\rho)}{\ff}+\mathcal O(1/\scrq^2)\,.
 \labell{einpot}
 \ee
Now this is precisely the effective potential which one would obtain
for Einstein gravity in the large momentum limit. There is an infinite
series of subleading corrections in $\mathcal O(1/\scrq^2)$ which
differ from Einstein theory but these terms are irrelevant in the limit
of large $\scrq$. This confirms our conclusion from the original
analysis of the effective dispersion relation.

Notice that in principle one has to worry that the boundary and large
momentum limits do not commute. This is clear from the form of
expression \reef{U exp}, where the momentum appears in ratios. We will
briefly comment on this in the discussion section. So perhaps a more
subtle analysis may still find new constraints from demanding causality
is respected in the dual CFT.

Our discussion here has focused on the tensor modes \reef{bang}.
However, the subtleties regarding the boundary and large $q$ limits
carry over to the vector and scalar channels. In the large $q$ limit,
one obtains the same results as in the tensor channel. That is, in this
limit the higher momentum terms dominate and one obtains a trivial
dispersion relation, leading to no causality violation. Now one may
wish to apply a more careful analysis for these modes, along the lines
of that given above for the tensor channel. However, unfortunately for
the vector and scalar channels, the previous analysis with an effective
Schr\"odinger equation can not be applied in a straightforward way
because of the appearance of higher powers of $\scrw$.

\section{Holographic hydrodynamics} \label{shear}

In this section, we compute the ratio of the shear viscosity to entropy
density for five-dimensional quasi-topological gravity. By now, the
holographic calculation of the shear viscosity is well understood. The
first computations of this transport coefficient from an AdS/CFT
perspective appeared in \cite{hydro} for Einstein gravity. These
calculations were soon after extended to include higher curvature
corrections to Einstein gravity, the first of example being the
computation of the leading corrections to $\eta/s$ for the strongly
coupled $\mathcal N=4$ super-Yang-Mills theory
\cite{BuchelDependence,buchel2,etas}. These computations were carried
out for GB gravity \cite{shenker1,GBanyd,spain1,GBrefs} and also higher
order Lovelock theories \cite{LLrefs}, where the higher derivative
terms need not be treated as small corrections. Further investigations
also provided increasingly efficient techniques for these calculations
\cite{LiuMembrane,newwald} In the following, we will use the `pole
method' of \cite{newwald}.

We begin with the metric for the planar AdS black hole given in
eq.~\reef{metric0} and which we write out again here
 \beq
\mathrm{d}s^2 = \frac{r^2}{L^2}\left(- \frac{f(r)}{\fin}
\,\mathrm{d}t^2 +\mathrm{d}x^2+\mathrm{d}x_2^2+\mathrm{d}x_3^2\right)
+\frac{L^2}{r^2 f(r)}\, \mathrm{d}r^2\,. \labell{metric0a}
 \eeq
Recall that $f(r)$ is determined by roots of the cubic equation in
eq.~\reef{constr1}. Now it is convenient to transform to a radial
coordinate $z=1-r^2_0/r^2$, with which the horizon is positioned at
$z=0$ and the asymptotic boundary, at $z=1$. The metric then becomes
 \beq
\mathrm{d}s^2 = \frac{r_0^2}{L^2(1-z)}\left(- \frac{f(z)}{\fin}
\,\mathrm{d}t^2 +\mathrm{d}x_1^2+\mathrm{d}x_2^2+\mathrm{d}x_3^2\right)
+\frac{L^2}{4 f(z)}\, \frac{\mathrm{d}z^2}{(1-z)^2}\,.
\labell{metric0b}
 \eeq
An important feature of these coordinates is that $f(z)$ has a simple
zero at the horizon. A Taylor expansion around $z=0$ yields
 \beq
f(z)= f'_0\,z+\frac 12\,f''_0\,z^2+\frac 16\,f'''_0\,z^3+\cdots\,,
 \labell{Taylor}
 \eeq
where, \eg $f'_0=\partial_zf|_{z=0}$. We present eq.~\reef{Taylor} to
establish a useful notation for the following.

Following \cite{newwald}, we perturb the metric \reef{metric0b} by
shifting
 \be
dx\to dx+\varepsilon\, e^{-i \omega t}\, dy \,, \labell{shift}
 \ee
where $\varepsilon$ is treated as an infinitesimal parameter. Then we
evaluate the Lagrangian density, \ie the entire integrand in
eq.~\reef{action} including $\sqrt{-g}$, on the shifted background to
quadratic order in $\varepsilon$. The presence of this off-shell
perturbation \reef{shift} produces a pole at $z=0$ in the (otherwise)
on-shell action. The shear viscosity is then given by the `time'
formula \cite{newwald}
 \be
\eta=-8\pi T \lim_{\omega,\varepsilon\to 0} \frac{\mbox{Res}_{z=0}\,
\mathcal L}{\omega^2\, \varepsilon^2}\,,
 \labell{shear0}
 \ee
where $\mbox{Res}_{z=0}\, \mathcal L$ denotes the residue of the pole
in the Lagrangian density. Recall the Hawking temperature for the above
black hole metric \reef{metric0b} is given in eq.~\reef{planarTh}. The
final result of this calculation for quasi-topological gravity is
 \beq
\eta=\frac{r_0^3}{2\lp^3 L^3}\Big[1-2 \lambda f'_0-9\mu\left(f'^2_0+
2 f''^2_0 +2 f'_0\left(f'''_0-3 f''_0 \right)\right)\Big]\,.
\labell{neweta}
 \eeq
Now with the $z$ coordinate, the cubic equation determining $f$ can be
written
 \be
f(z)-\lambda f(z)^2-\mu f(z)^3=z\,(2-z)
 \labell{cubic9}
 \ee
and substituting in the Taylor expansion \reef{Taylor}, we can
explicitly determine the coefficients:
 \be
 f'_0=2\,,\qquad f''_0=-2(1-
4\lambda)\,,\qquad f'''_0=-24(\lambda -4\lambda^2 -2\mu)\,.
 \labell{fzero}
 \ee
The above expression \reef{neweta} for the shear viscosity then becomes
 \be
\eta=\frac{r_0^3}{2\lp^3}\big[ 1-4\lambda-36 \mu(9-64 \lambda+128
\lambda^2+48 \mu)\big]\,. \labell{neweta2}
 \ee
We readily verify that with $\mu=0$, this result \reef{neweta2} reduces
to the expected result for GB gravity \cite{shenker1}.

\FIGURE{ \centering
	\includegraphics[width=120mm]{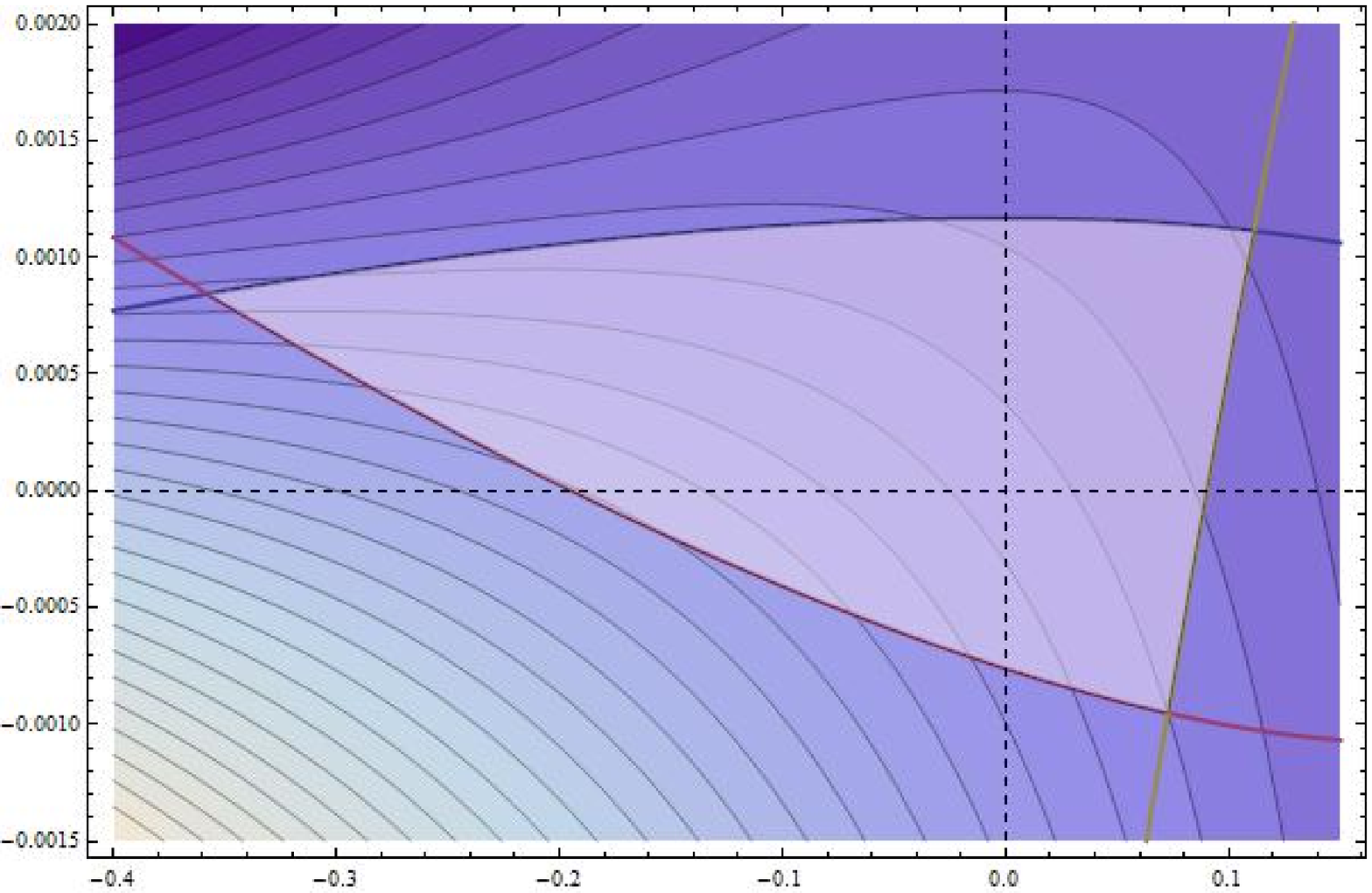} \caption{Contours of
constant $\eta/s$ shown in the allowed region of the gravitational
couplings --- see figure \ref{3ptplot}. The ratio increases in going
from the lower-left to the upper-right in the plot. } \labell{etaplot}
}
Combining this result with eq.~\reef{entropy}, we find
 \be
\frac{\eta}{s}=\frac{1}{4 \pi}\left[ 1-4\lambda-36 \mu(9-64 \lambda+128
\lambda^2+48 \mu)\right]\, .\labell{etas}
 \ee
A contour plot of the ratio of the shear viscosity to the entropy
density in space of gravitational couplings, $\lambda$ and $\mu$, is
shown in figure \ref{etaplot}. From this plot, it is evident that
$\eta/s$ is maximized in the upper-right corner of the the allowed
region of couplings. This point corresponds to the intersection of the
boundaries defined by the tensor and vector constraints, \ie
eqs.~\reef{spin2} and \reef{spin1}, respectively. At this point, one
finds
 \be
\fin=\frac{2838}{2543}\,,\quad \lambda=\frac{246671}{2684748} \,, \quad
\mu=\frac{6466849}{5714486118}\,. \labell{corner}
 \ee
Hence we find the minimum value for $\eta/s$ for the class of
four-dimensional CFT's dual to quasi-topological gravity is
 \be
\left.\frac{\eta}{s}\right|_{min}=\frac{347182615788747017}
{838580510094780681}\, \frac 1{4\pi} \simeq (.4140)\,\frac 1{4\pi}\, .
 \labell{min}
 \ee
Note that this point is well away from the region of instabilities
which will be discussed in the next section. Hence we expect that our
calculation of $\eta/s$ at this point is reliable.

Within this class of CFT's, there is also a maximum value for $\eta/s$
which appears to occur near the midpoint of the boundary produced by
the scalar constraint \reef{spin0}. However, the point of the precise
minimum lies in a region where, in the next section, we find that the
uniform plasma is unstable --- see figure \ref{Instability-region}.
Hence our hydrodynamic calculations are not reliable at this precise
point. Excluding the unstable region, it appears the maximum occurs
very close to the point in GB gravity where $\eta/s$ is maximized
\cite{GBanyd,spain1}. That is, the maximum is near
$(\lambda,\mu)=(-1/8,0)$ where we find
 \be
\left.\frac{\eta}{s}\right|_{max}\simeq\frac 32\,\frac 1{4\pi}\, .
 \labell{max}
 \ee

\section{Plasma Instabilities} \label{plastab}

Even when the various consistency conditions of section \ref{harmsway}
are satisfied, there remains the possibility that the black hole
solution is unstable. The dual statement would be that an infinite
uniform plasma is an unstable configuration for the CFT. Such an
instability need not represent a fundamental pathology of the CFT but
rather indicate that some interesting new dynamics arises in CFT plasma
for certain values of the couplings. However, it is still important to
identify such instabilities as they would invalidate the assumption of
local thermodynamic equilibrium and for example, discredit the results
of our hydrodynamic calculations in section \ref{shear}.

The appearance of such instabilities for five-dimensional GB theory
were first noted in \cite{shenker1,alex0}. Although causality or
positive flux constraints allow the GB coupling to be in the range
$-7/36 \le \lambda\le 9/100$, one finds that for $\lambda<-1/8$ a new
instability arises. Evidence for the latter was given as follows: First
one writes the equation of motion for the tensor modes in an effective
Schr\"odinger form, as was done in section \ref{causalconstr} above.
For $\lambda<-1/8$, the potential develops a small well where $U<0$
just in front of the horizon (\ie near $\rho=1$). For sufficiently
large $\scrq$, this well will support negative energy bound states
which then correspond to unstable quasinormal modes, as described in
\cite{andrei}. Given that $\scrq$ is finite (and large), this
instability indicates that the uniform plasma becomes unstable with
respect to certain non-uniform perturbations. On the gravitational side
then, this instability seems similar in certain respects to the
Gregory-Laflamme instability for black strings \cite{gl}. However,
while the latter involves long wavelength modes, here the `plasma
instatiblity' occurs for arbitrarily short wavelengths. Examining the
sound and shear channels, one finds that no additional instabilities
arise in the consistent range, $-7/36 \le \lambda\le 9/100$
\cite{alex0}. The same analysis has also been extended to GB gravity in
higher dimensions \cite{GBanyd}. For $D=6$, one finds similar range
where the theory passes all the known consistency tests but the uniform
plasma is unstable. However, for $D\ge7$, all of the potential
instabilities are pushed outside of the allowed range of the GB
coupling.

In this section, we will provide a preliminary investigation of
potential plasma instabilities for five-dimensional quasi-topological
gravity. Following the discussion above, our strategy will be to
examine the tensor modes in detail using the effective Schr\"odinger
equation, which was presented in eq.~\reef{schro99}. As described
previously, because of the appearance of higher powers of $\scrw$ in
the sound and shear mode equations, one cannot construct an effective
Schr\"odinger problem in these cases. Hence a more elaborate analysis
of the quasi-normal modes would be required to detect instabilities in
these channels.

We will separate our analysis into several different regimes, as we
will find the behaviour of the theory will be quite different depending
on the the sign of $\mu$ and the magnitude of $\scrq^2$. To see this,
consider eq.~\reef{PreSchro}. In particular, let us examine the
coefficients $A(\rho,\scrq^2)$ and $D(\rho,\scrq^2)$:
\bea
A(\rho,\scrq^2)&=& -\rho^2 \frac{d}{d\rho}\bigg[ \frac 1\rho
\left(1-2 \la f(\rho)-3 \mu f(\rho)^2\right) \labell{bigA}\\
&&\qquad+ 9 \mu \,\rho \, \frac d{d\rho} \left(\rho f'(\rho)^2\right) -
12 \mu \frac{\scrq^2}{\ff} \left(f(\rho)-2 \rho f'(\rho)\right)
\bigg]\,,
 \nonumber\\
D(\rho,\scrq^2)&=&\frac{A(\rho,\scrq^2)}{\rho\, f(\rho)^2}-9 \mu
\frac{\rho}{f(\rho)} \left( 9 f'(\rho)+16 \rho\,
f^{(3)}(\rho)+4\rho^2\, f^{(4)}(\rho)\right)\,. \labell{bigD}
 \eea
The presence of the term proportional to $\mu\,\scrq^2$ in the first
expression creates the possibility that $A$ (or $D$) may vanish, which
we will see leads to a singularity in the Schr\"odinger potential. To
see how this zero comes about, we first evaluate these functions at the
horizon (\ie $\rho=1$). Using eq.~\reef{constr1}, the polynomial
defining $f(\rho)$, and $\rho=r_0^2/r^2$, we find
 \be
f(\rho)\simeq -2 (\rho-1)-(1-4\lambda) (\rho-1)^2+\cdots\,.
 \labell{fhorz}
 \ee
With this result, we find the corresponding expansions of $A$ and $D$
at the horizon
 \be
A\simeq A_0 +\mathcal O(\rho-1)\,, \qquad
D = \frac{A_0}{4(\rho-1)^2}+\mathcal O(\rho-1) \labell{expunge}
 \ee
with
 \be
 A_0= 24(3-8 \lambda) \mu\,
\frac{\scrq^2}{\ff}+\left[1-4\lambda-36\mu(9-64\lambda+128
\lambda^2+48\mu)\right]\,. \labell{a0}
 \ee
Hence we see that $A$ vanishes at the horizon for a specific critical
value of $\scrq^2$:
 \be
\scrq^2_c=\frac{1-4\lambda-36\mu(9-64\lambda+128
\lambda^2+48\mu)}{24(3-8 \lambda)(-\mu)}\,.
 \labell{qcrit}
 \ee
Now comparing with eq.~\reef{etas}, we see the expression in the
numerator above is precisely $4 \pi \eta/s$. In the physically allowed
region found in section \ref{constraints}, this ratio is always
positive, as is the factor $3-8\lambda$ --- see figure \ref{3ptplot}.
Therefore we only have a valid solution for $\scrq_c$ when $\mu$ is
negative.

Note that $A_0$ vanishes when $\scrq=\scrq_c$ and so from
eq.~\reef{expunge}, both $A$ and $D$ vanish on the horizon at this
point. For larger values of $\scrq$, both $A$ and $D$ have a zero
outside of the horizon (\ie $\rho<1$) but the two zeros appear at
different radii.

From the above discussion, it is clear that we should separate the
analysis into three distinct regimes: i) $\mu>0$, ii) $\mu<0$ and
$|\scrq|\le\scrq_c$ and iii) $\mu<0$ and $|\scrq|>\scrq_c$.

\subsection{Positive $\mu$} \label{bug2}

In the case where $\mu\geq 0$, the functions $A$ and $D$ are positive
everywhere outside of the horizon. Hence, the effective Schr\"odinger
potential \reef{potu1} is well behaved everywhere in the range of
interest, $0\le\rho\le1$. To identify instabilities, it may seem that
we can apply directly the strategy described above for GB theory of
looking for a small negative dip in the effective potential just
outside the horizon \cite{shenker1,alex0,GBanyd}. However, there is a
small subtlety which requires a more detailed investigation for
quasi-topological gravity. For GB gravity, one further considers the
limit of large $\scrq$, which corresponds to the limit of $\hbar\to 0$
from the point of view of the effective Schr\"odinger problem. With
this limit, any small negative dip will always lead to negative energy
bound states as solutions to eq.~\reef{schro99} and hence the
appearance of an instability \cite{andrei}. However, as found in
section \ref{causalconstr}, the structure of the potential strongly
depend on the value of the momentum for quasi-topological gravity.
Indeed, our analysis showed that for sufficiently large momentum the
effective potential reduces to that of Einstein gravity, \ie $U\simeq
f(\rho)/\fin$, and so no unstable modes would appear in this limit.
That is, if one starts near an unstable point in GB theory but now with
a small positive $\mu$, then generically the modes become more stable
as the momentum is increased. Hence in general one will have to
investigate the potential for finite values of $\scrq$ to find any
unstable modes.

Following the reasoning of \cite{andrei}, we will still identify the
unstable quasinormal modes as negative energy bound states in the
effective Schr\"odinger problem. For a moment, let us consider applying
the Bohr-Sommerfeld quantization rule for a zero-energy bound state:
 \be \scrq \int^{1}_{\rho_0} d\rho\,\frac{dy}{d\rho}\,\sqrt{-U(\rho,
\scrq^2)}=\left(n-\frac 12 \right) \pi \labell{bs int}\,.
 \ee
Here, we are assuming that the potential dips to below zero between the
horizon at $\rho=1$ and some lower turning point where $U(\rho=\rho_0,
\scrq^2)=0$. For the zero-energy state, the quantum number $n$ would be
a specific positive integer. More generally, if we were to evaluate the
integral on the left-hand side, then the integer part of $n$ on the
right would count the number of negative energy bound states supported
by the potential well. Hence our strategy here is to scan of the
parameter space $(\lambda,\mu)$ for positive $\mu$. At each point, we
vary $\scrq$ looking for a negative dip in the potential near horizon.
If the dip becomes sufficiently deep to support a bound state (\ie to
satisfy eq.~\reef{bs int} with $n=1$), we will take this to signal of
an instability in the system.

As discussed above for GB gravity, it was found that instabilities
appear for $\lambda\leq -1/8$. Hence we focussed our scan of parameters
on small positive values of $\mu$ in this regime and our numerical
results are presented in figure \ref{bs}. We should explicitly say that
we expect that our Bohr-Sommerfeld analysis gives a good guide as to
the unstable parameter space but it is difficult to assess how accurate
the boundary of this region is in figure \ref{bs}. In this regard, it
is reassuring that the numerical curve reaches the $\lambda$ axis very
close to $\lambda=-0.125$ which, as indicated above, is where previous
analysis \cite{shenker1,alex0} indicated that GB gravity should become
unstable.
\FIGURE{ \centering
	\includegraphics[width=12 cm]{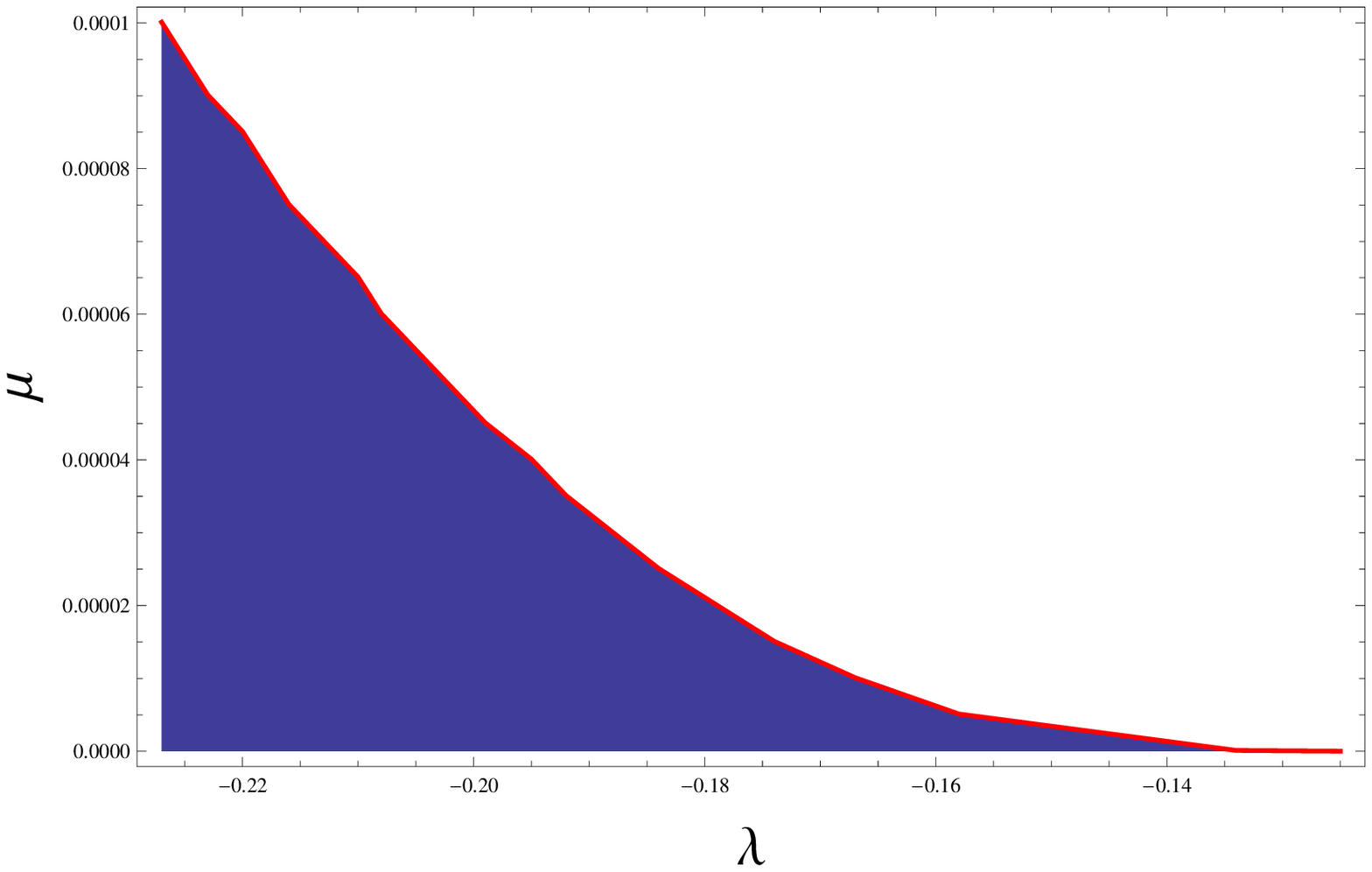} \caption{Instability boundary
for the positive $\mu$ estimated using the Bohr-Sommerfeld analysis.
The (blue) region below the (red) curve is unstable.}
	\labell{bs} }

Note that the unstable region is very narrow just above the $\lambda$
axis but the height of this region increases as $\lambda$ becomes more
negative. Intuitively, this behaviour arises because for GB gravity at
$\mu=0$, an infinitesimal negative dip first appears in the potential
at $\lambda=-1/8$ and then becomes larger as $\lambda$ is decreased
further. An infinitesimal dip appears is sufficient to support negative
energy states in GB gravity because $\scrq$ can be taken arbitrarily
large without effecting the effective potential. As discussed above,
with nonvanishing $\mu$, increasing the momentum makes the potential
more stable or, in other words, decreases the size of dip in $U$. Hence
one must balance this effect with the increase in the pre-factor
$\scrq$ in eq.~\reef{bs int}. Therefore it is easier to find an
instability for positive $\mu$ when when the initial size of the dip is
larger at the corresponding point on the $\lambda$ axis.

\subsection{Negative $\mu$ and $|\scrq|\le\scrq_c$} \label{bug1}

From our introductory discussion, we expect that the structure of the
potential may be radically different for negative $\mu$ and
particularly for $|\scrq|\ge\scrq_c$. Here we will examine the approach
to the critical momentum, $\scrq\rightarrow \scrq_c$, and the same
Bohr-Sommerfeld analysis as above will show that unstable modes occur
over a large part of this parameter regime.

We are again looking for a negative potential well in front of the
horizon, now for $\mu<0$ and $|\scrq|\le\scrq_c$. So to begin, consider
the near horizon expansion of the effective potential \reef{potu1}:
\be U(\rho, \scrq^2)\simeq \frac{U_0}{A_0}\,
(\rho-1)+\frac{U_1}{(A_0)^2}\,(\rho-1)^2+\cdots\,,
 \labell{nearU}
 \ee
where $U_0$ and $U_1$ are some constants and $A_0$ is given in
eq.~\reef{a0}. An important point to notice is that as the momentum
increases towards the critical value, $\scrq^2\rightarrow\scrq_c^2$, we
have $A_0\ll 1$ and the above expansion becomes ill-defined. One must
really perform a separate expansion for the special case
$\scrq^2=\scrq_c^2$, which yields
 \be
U(\rho, \scrq_c^2)\simeq \frac 1{\scrq_c^2}
\,\frac{P(\lambda,\mu)}{Q(\lambda,\mu)}+\mathcal O(\rho-1)\,.
 \labell{special}
 \ee
where $P$ and $Q$ are polynomials in the couplings $\lambda$ and $\mu$.
Their details are not important but for completeness we present them
here
 \bea
P&=& 2 (1-4 \lambda )^2 \lambda +3 (-1+4 \lambda ) \left(-41+32 \lambda
   \left(19-76 \lambda +96 \lambda ^2\right)\right) \mu\nonumber \\
   && \qquad -36 (1393+12
   \lambda  (-1905+8 \lambda  (1469+176 \lambda  (-23+24 \lambda ))))
   \mu ^2 \nonumber \\
   &&\qquad - 5184 (161+24 \lambda  (-49+96 \lambda )) \mu ^3-1990656 \mu
   ^4\,, \labell{PPP} \\
Q&=& 3 (-1+4 \lambda ) (9+8 \lambda  (-5+8 \lambda )) \mu -288
\left(20+3
   \lambda  \left(-189+16 \lambda  \left(97-310 \lambda +336 \lambda
   ^2\right)\right)\right) \mu ^2 \nonumber \\
   &&\qquad - 10368 (-1+4 \lambda ) (-45+104
   \lambda ) \mu ^3+248832 \mu ^4\,.
   \labell{QQQ}
 \eea
The essential result here is that while for $\scrq^2<\scrq_c^2$, the
effective potential went to zero on the horizon, precisely at the
critical momentum, it is tending to a constant value at the horizon.
Hence the limit $\scrq^2\rightarrow\scrq_c^2$ is actually
discontinuous. This unusual behaviour is illustrated with an example in
figure \ref{triangle}.
\FIGURE{ \centering
	\includegraphics[width=12cm]{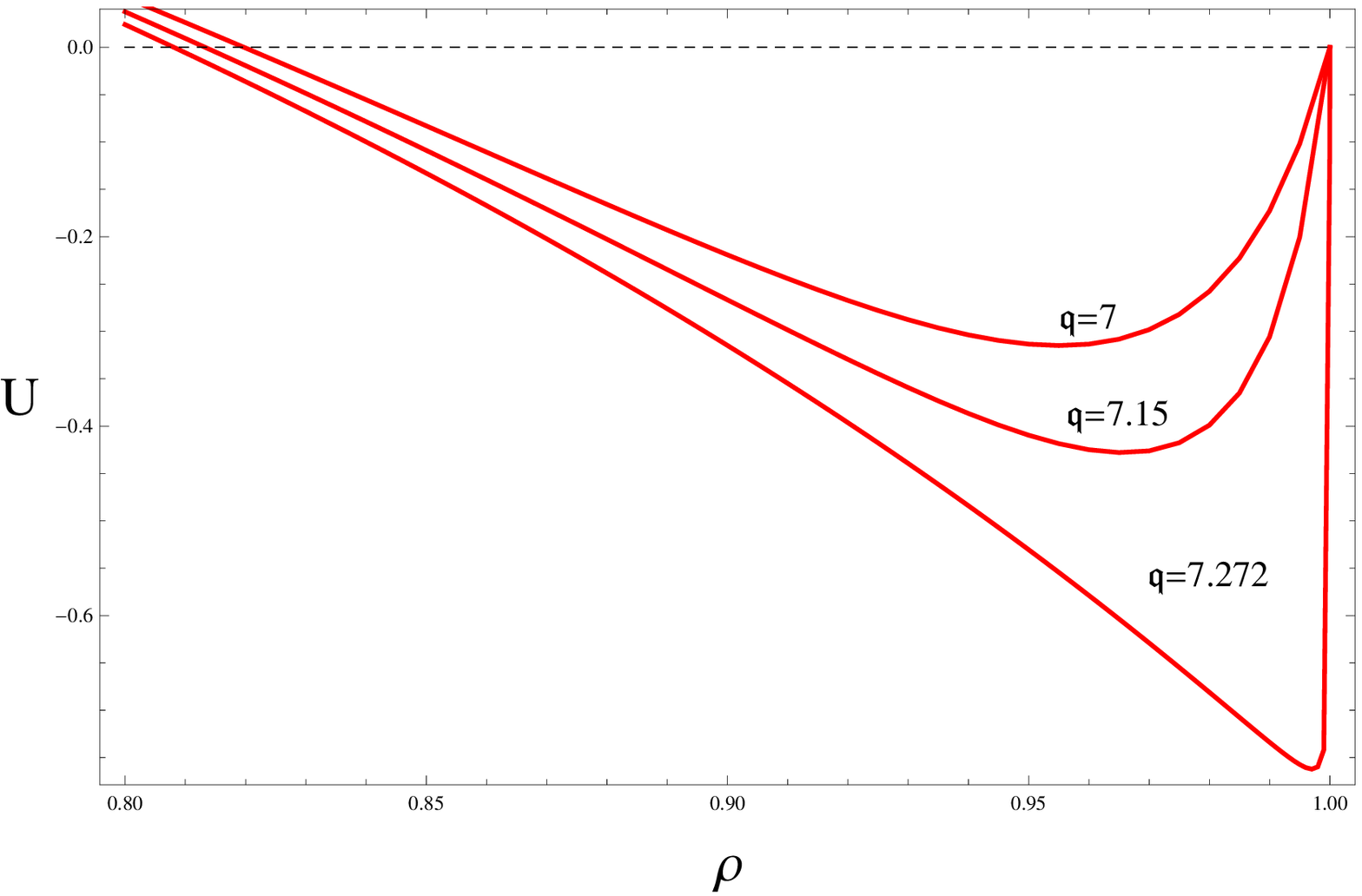} \caption{ The effective
Schr\"odinger potential in the tensor channel for $\mu=-0.0003,
\lambda=-0.1$ and different values of $\scrq^2$. Here $\scrq_c\simeq
7.2725$. The effective potential at the horizon is tending to the
constant value $P/(\scrq_c^2 Q)\simeq -0.797$}
	\labell{triangle} }

The example in this figure also illustrates if this limiting constant
value is negative, \ie $U(\rho=1,\scrq^2=\scrq_c^2)<0$, then the
effective potential develops a negative well in front of the horizon as
we approach the critical momentum. Hence there is the possibility of
developing instabilities in this regime. The dip in potential is
largest with $\scrq^2=\scrq^2_c$ and so we focus on this case. Once
again, we will use the Bohr-Sommerfeld rule \reef{bs int} to test for
negative energy bound states. As before, we also have to worry that
$A_0\to 0$ as $\scrq^2\to \scrq^2_c$ and so we must perform a separate
expansion to evaluate $dy/d\rho$ at this point
 \be
\frac{dy}{d\rho}\simeq -\frac {j_0}{2(\rho-1)}+\cdots\,, \qquad {\rm
where}\ j_0= \sqrt{\frac{3 \mu\, Q(\lambda,\mu)}{R(\lambda,\mu)}}\,,
 \ee
where $Q$ was given above in eq.~\reef{QQQ} and
 \be
R= 3 \mu Q+108 \,\mu \, (1-4\lambda)(3-8\lambda)\left(3-80
\lambda(1-4\lambda)+320\mu\right)\,.
 \labell{RRR}
 \ee
Hence the Bohr-Sommerfeld integral \reef{bs int} yields an expression
of the form
 \be
\left(n-\frac 12 \right) \pi\simeq\frac{1}2 \int_1^{\rho_0} d\rho
\left(\frac{1}{\rho-1} \sqrt{\frac{-3\mu\,P}{R}}+\cdots\right)\,.
 \labell{diverge2}
 \ee
This integral produces logarithmic divergence indicating that there are
an infinite number of unstable modes in this limit. Of course, this
result is only true at strictly $\scrq=\scrq_c$. However, by taking the
momentum arbitrarily close to the critical value, the resulting
negative dip in the potential will always support a large number of
unstable modes as well.

To summarize, we must exclude as unstable the region with $\mu<0$ where
the effective potential tends to a negative constant on the horizon at
the critical value $\scrq_c$. That is, the region where $\mu<0$ and
$P/Q\leq 0$. The boundary of this region is well approximated by the
curve
\be \mu\simeq
-\frac{2}{123}\,\lambda-\frac{14120}{206763}
\,\lambda^2-\frac{27472807424}{28153056843}\,\lambda^3.
 \labell{boundinst}
 \ee
Putting this constraint for negative $\mu$ together with that obtained
for positive $\mu$ in the previous section, one obtains the unstable
region shown in figure \ref{Instability-region}. Here we are showing
the region of instability together with the physically allowed region
defined by the constraints in eqs.~(\ref{spin2}--\ref{spin0}).
\FIGURE{ \centering
	\includegraphics[width=12 cm]{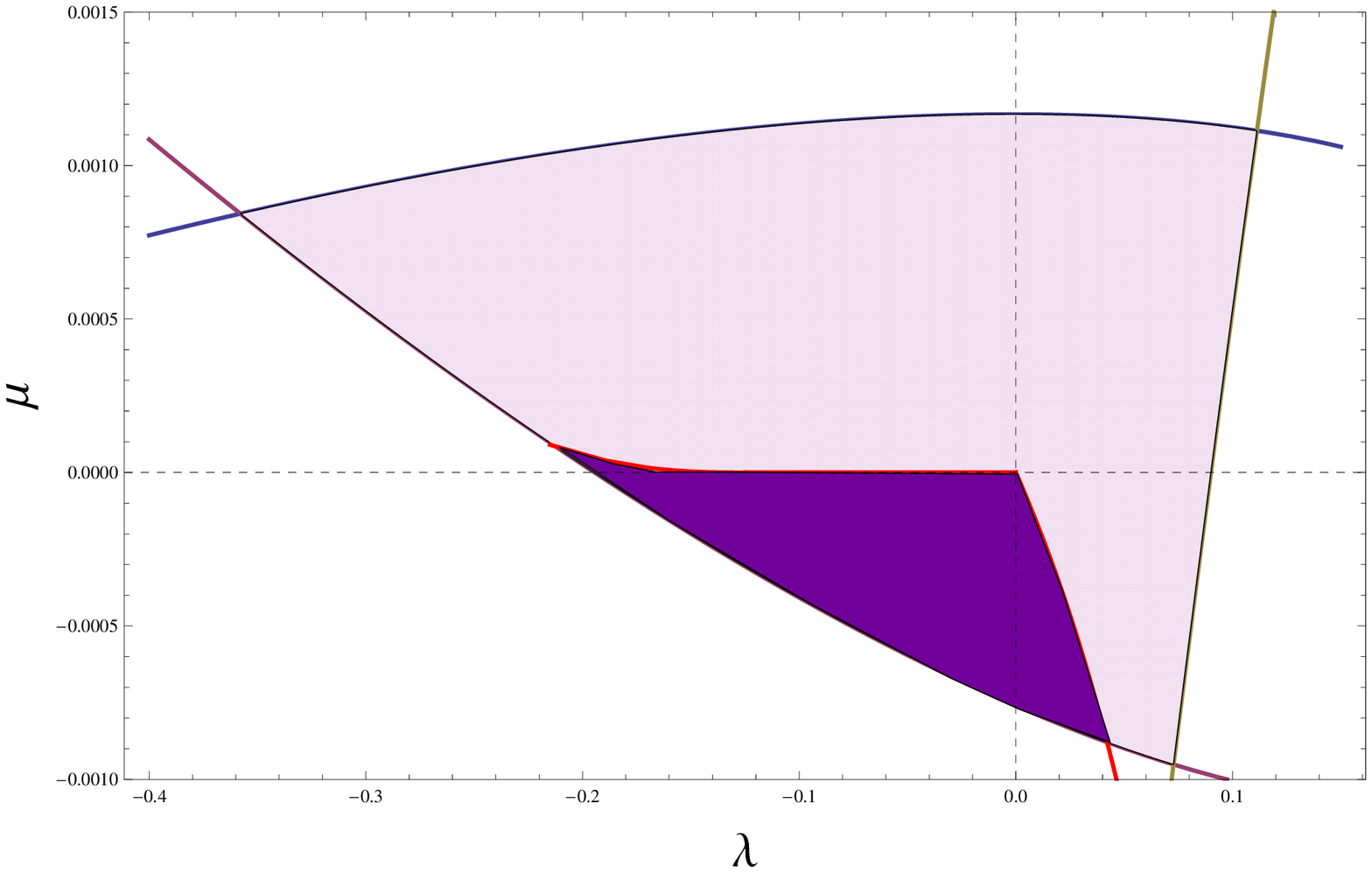}
\caption{Unstable region (purple) for the tensor channel combining the
results of sections \ref{bug2} and \ref{bug1}, superimposed onto the
region (pink) allowed by positivity of energy flux. In the section
\ref{bug}, the analysis is extended to $|\scrq|>\scrq_c$ and it appears
instabilities are present throughout the entire $\mu<0$ region.}	
\labell{Instability-region} }

\subsection{Negative $\mu$ and $|\scrq|>\scrq_c$} \label{bug}

With $\mu<0$, we are of course free to take the momentum beyond the
critical value. As noted above, when $\scrq^2>\scrq_c^2$, both $A$ and
$D$ have a zero at some finite radius outside of the horizon but the
two zeros appear at different radii. Let us label the two zeros as
$\rho_A$ and $\rho_D$. In the following, we consider the case where
$\rho_A>\rho_D$, but in general this depends in detail on the values of
$\lambda$ and $\mu$. However, the following analysis is easily adapted
to the opposite situation where $\rho_A<\rho_D$ and the conclusions
will be unchanged.

Let us first consider the effective potential in the vicinity of $D$'s
zero. Expanding the various coefficients in eq.~\reef{PreSchro} around
$\rho\simeq\rho_D$ gives
 \bea
A&=& A_0+\mathcal O\left(\rho-\rho_D\right)\,, \nonumber \\
B&=& B_0+\mathcal O\left(\rho-\rho_D\right)\,, \nonumber \\
C&=& C_0+\mathcal O\left(\rho-\rho_D\right)\,, \nonumber \\
D&=& -D_1(\rho-\rho_D)+\mathcal O\left((\rho-\rho_D)^2\right)\,.
 \eea
It is important to note that the constants $A_0$ and $D_1$ are both
positive, so that $dy/d\rho$ in eq.~\reef{useful2} is well defined in
the region $\rho<\rho_D$. Using the expressions above, along with
eq.~\reef{useful2}, we find
\bea
y_D-y&\simeq& \frac 23 \sqrt{\frac{D_1}{A_0}} \,(\rho_D-\rho)^{3/2}\,
 \labell{Done} \\
Z&\simeq&(\rho_D-\rho)^{-1/4}\,. \label{Dtwo}
 \eea
Finally, with eq.~\reef{potu1}, we obtain the effective potential
 \bea
\scrq^2\, U(\rho)&=& -\frac{A_0}{D_1}\, \frac{5/16}{(\rho-\rho_D)^3} \nonumber \\
&=&-\frac{5/36}{(y_D-y)^2}\,.\labell{Dpot}
 \eea
Hence we find that the effective potential contains a singularity at
$\rho=\rho_D$ but the structure of the singularity is surprisingly
simple. In particular, the coefficient is completely independent of the
parameters, $\lambda$, $\mu$ and $\scrq$, but of course, the latter
still determine the precise location of the singularity.

We can perform a similar analysis for the effective potential in the
vicinity of the zero in $A$. We begin by expanding around
$\rho\simeq\rho_A$
\bea
A&=& -A_1\left (\rho-\rho_A\right ) +\mathcal O\left((\rho-\rho_D)^2\right)\,,\nonumber\\
B&=& -A_1+\mathcal O\left(\rho-\rho_A\right)\,,\nonumber\\
C&=& \hat C_0+\mathcal O\left(\rho-\rho_A\right)\,,\nonumber\\
D&=&-D_0+\mathcal O\left(\rho-\rho_A\right)\,.
 \eea
for some positive constants $D_0$ and $A_1$. Note that the leading
coefficient in $B$ has the special form $B_0=A_1$, which follows from
the original equation \reef{genZ0}.  Then  in this case
 \bea
y-y_A&\simeq& 2 \sqrt{\frac{D_0}{A_1}} \,(\rho-\rho_A)^{1/2}\,,
 \labell{Aone} \\
Z&\simeq&(\rho-\rho_A)^{-1/4} \labell{Atwo}
 \eea
and the potential becomes
\bea \scrq^2\, U(\rho)&=& -\frac{A_1}{D_0}\, \frac{1/16}{\rho-\rho_A}
 \nonumber \\
&=&-\frac{1/4}{(y-y_A)^2}\,.
 \labell{Apot}
 \eea
Here again we find that the effective potential contains a singularity
with a very simple form at the zero of $A$. The parameters, $\lambda$,
$\mu$ and $\scrq$, fix the precise location of the singularity but the
overall coefficient is independent of these.

Implicitly the above analysis assumes that $\rho>\rho_A$ where both $A$
and $D$ are negative. Similarly in consider the zero in $D$, we assumed
$\rho<\rho_D$ where both $A$ and $D$ are positive. Special
consideration must be given to the range $\rho_D<\rho<\rho_A$ where
$D<0$ and $A>0$. To accommodate the latter, we must modify the
construction of the effective Schr\"odinger equation \reef{schro99}
slightly. In particular, in eq.~\reef{useful2}, the definition of the
new coordinate is replaced by
 \be
\frac{dy}{d\rho}= \sqrt{-\frac{D(\rho,\scrq^2)}{A(\rho,\scrq^2)}}\,.
\labell{useful3x}
 \ee
The final result can then be most simply expressed as flipping the sign
of both the effective potential and the effective energy of
eq.~\reef{schro99}. That is, in this region, the Schr\"odinger equation
becomes
 \be
-\frac{1}{\scrq^2}\, \partial_y^2 \psi(y)+ \left[-U(y,\scrq^2)\right]\,
\psi(y)=\left[-\alpha^2\right]\,\psi(y)\,,\labell{schro99a}
 \ee
where $U$ is given by precisely the same expression in eq.~\reef{potu1}
and $\alpha^2=\scrw^2/\scrq^2$, as before. For present purposes, the
behaviour near the zeros $\rho=\rho_D$ and $\rho_A$ are of primary
interest. Carefully keeping track of the signs, one finds that near
these points, the Schr\"odinger equation \reef{schro99a} has precisely
the same form as above, with a singular and attractive potential. That
is, in the vicinity of these zeros, we may write the Schr\"odinger
equation as
 \be
-\psi''(y)-\frac{\gamma}{y^2}\, \psi(y)\simeq 0
 \labell{schror2}
 \ee
where $\gamma=5/36$ for $\rho=\rho_D$ and $1/4$ for $\rho=\rho_A$. In
either case we have shifted the $y$ coordinate to put the singularity
at the origin. The $1/y^2$ potential has been studied extensively in
the literature \cite{r2papers}. Remarkably for a one-dimensional
Schr\"odinger equation, the value $\gamma=1/4$ marks the boundary
between the conformal regime when the potential is repulsive or weakly
attractive and the regime with a sufficiently attractive potential
where discrete bound states appear and conformality is lost.

The key point in presenting eq.~\reef{schror2} that this equation
applies here for both positive and negative $y$. However, in this
one-dimensional setting, the wave-function must propagate through the
singularity at $y=0$. So one should approach the problem by solving for
$y<0$ and $y>0$ independently and then matching the solutions with an
appropriate boundary condition at the origin. This procedure is most
easily demonstrated for the zero in $D$, in which case $\gamma=5/36$.
Then eq.~\reef{schror2} has two independent solutions
 \bea
\psi(y)&=& d_1 \, y^{1/6}+d_2 \, y^{5/6} \qquad \qquad \qquad\quad{\rm
for}\ \ y>0\,,
  \nonumber \\
\psi(y)&=& \tilde d_1 \, (-y)^{1/6}+ \tilde d_2 \, (-y)^{5/6} \qquad
\qquad {\rm for}\ \ y<0\,. \nonumber
 \eea
Using eqs.~\reef{Done} and \reef{Dtwo} above, we can convert these back
to the radial profile of the original tensor mode, \ie $\phi=Z \psi$,
in the vicinity $\rho\simeq\rho_D$
 \bea
\phi(\rho)&= & \delta_1 +\delta_2 (\rho-\rho_D) \qquad \qquad {\rm
for}\ \ \rho<\rho_D\,,
 \nonumber \\
&= & \tilde\delta_1 +\tilde\delta_2 (\rho-\rho_D) \qquad \qquad {\rm
for}\ \ \rho>\rho_D\,.
 \eea
Matching of solutions at $\rho_D$ is achieved straightforwardly by
imposing continuity of the radial profile and its first derivative, \ie
$\delta_1=\tilde\delta_1$ and $\delta_2=\tilde\delta_2$. Hence while
some care must be taken, the zero in $D$ presents no real difficulty in
finding well-behaved solutions.

Next we turn to the zero in $A$, where the situation is more subtle. In
this case, $\gamma=1/4$ in eq.~\reef{schror2} and the two independent
solutions are
 \bea
\psi(y)&=& a_1 \, y^{1/2}+a_2 \, y^{1/2}\, \log(y) \qquad \qquad
\qquad\qquad{\rm for}\ \ y>0\,,
  \nonumber \\
\psi(y)&=& \tilde a_1 \, (-y)^{1/2}+ \tilde a_2 \, (-y)^{1/2}\,
\log(-y) \qquad \qquad {\rm for}\ \ y<0\,. \nonumber
 \eea
Using eqs.~\reef{Aone} and \reef{Atwo}, these expressions are
translated to the radial profile $\phi=Z \psi$ in the vicinity
$\rho\simeq\rho_A$
 \bea
\phi(\rho)&= & \alpha_1+\alpha_2 \log(\rho-\rho_A), \qquad \qquad {\rm
for}\ \ \rho>\rho_A\,,
 \nonumber \\
&= & \tilde\alpha_1+\tilde\alpha_2 \log(\rho_A-\rho),\qquad \qquad {\rm
for}\ \ \rho<\rho_A\,.
 \eea
The matching at $\rho_A$ is slightly more involved because of the
logarithmic behaviour of these solutions. Integrating the equation of
motion \reef{PreSchro} around $\rho_A$, we obtain
\be -A_1 (\rho-\rho_A) \partial_\rho \phi \big
|_{\rho_A-\epsilon}^{\rho_A+\epsilon}=\int_{\rho_A-\epsilon}^{\rho_A+\epsilon}
d\rho\, \left( D_0\,\scrw^2-\hat C_0\right) \phi\,.
 \ee
As $\epsilon\rightarrow0$, the right-hand side vanishes and the
vanishing of the left-hand side requires $\alpha_2+\tilde\alpha_2=0$.
Another natural boundary condition is that the flux of probability in
the effective Schr\"odinger problem should be continuous as the
wave-function propagates through the singularity at $y=0$. If we
express this flux in terms of the radial profile $\phi$ and the
coordinate $\rho$, then we require
$$
 \mbox{Im} \big[\phi^* A_1(\rho-\rho_A)\partial_\rho \phi
 \big]_{\rho_A-\epsilon}^{\rho_A+\epsilon}=0\,.
$$
Continuity then leads to $\mbox{Im}(\alpha_1^*
\alpha_2)=\mbox{Im}(\tilde \alpha_1^* \tilde \alpha_2)$ and so it seems
natural to set $\alpha_1=-\tilde\alpha_1$ as well.\footnote{This is not
a unique solution for this constraint and so it may be that a further
boundary condition should be applied to single out this result as the
unique physical solution.} Our final solution in the vicinity of
$\rho\simeq\rho_A$ then takes the form
 \be
\phi=\left(\alpha_1+\alpha_2 \log |\rho-\rho_A|\right)\,{\rm
sgn}(\rho-\rho_A)\,.
 \ee
Again the constants $\alpha_1$ and $\alpha_2$ are arbitrary but our
analysis shows appropriate boundary conditions at this point will
produce a suitable physical solution.

At this stage, we have shown that despite of the appearance of new
singularities in the equation of motion \reef{PreSchro} in this regime
(\ie $\mu<0$ and $|\scrq|>\scrq_c$), we may more or less
straightforwardly solve for the radial profile $\phi$. The precise
solutions and the corresponding quasinormal eigenfrequencies $\scrw$
will still be set by the boundary conditions on the field at the black
hole horizon $\rho=1$ and at the asymptotic boundary $\rho=0$. To
better understand these boundary conditions, we now return to the
overall behaviour of the effective Schr\"odinger potential.

Figure \ref{compare-beyond} provides an example of the effective
potential in the desired regime. One point which the figure illustrates
is that for large momenta $\scrq^2\gg\scrq^2_c$, the general discussion
of section \ref{causalconstr} still applies here and over most of the
range the effective potential approaches that of Einstein gravity, \ie
$U\simeq f(\rho)/\fin$. However, the figure also makes evident the
singularities extensively discussed above, which appear as sharp dip at
$\rho=\rho_A$ and $\rho_D$. As is typical of $\scrq^2\gg\scrq^2_c$, the
zeros are very close together and in fact it is difficult to resolve
the two distinct singularities in the example given in figure
\ref{compare-beyond}. Intuitively, one expects that this deviation of
the Einstein potential will provide a small perturbation and so there
will be a set of stable modes whose quasinormal frequencies and radial
profiles deviate only slightly from the solutions in Einstein gravity.
\FIGURE{ \centering
	\includegraphics[width=12 cm]{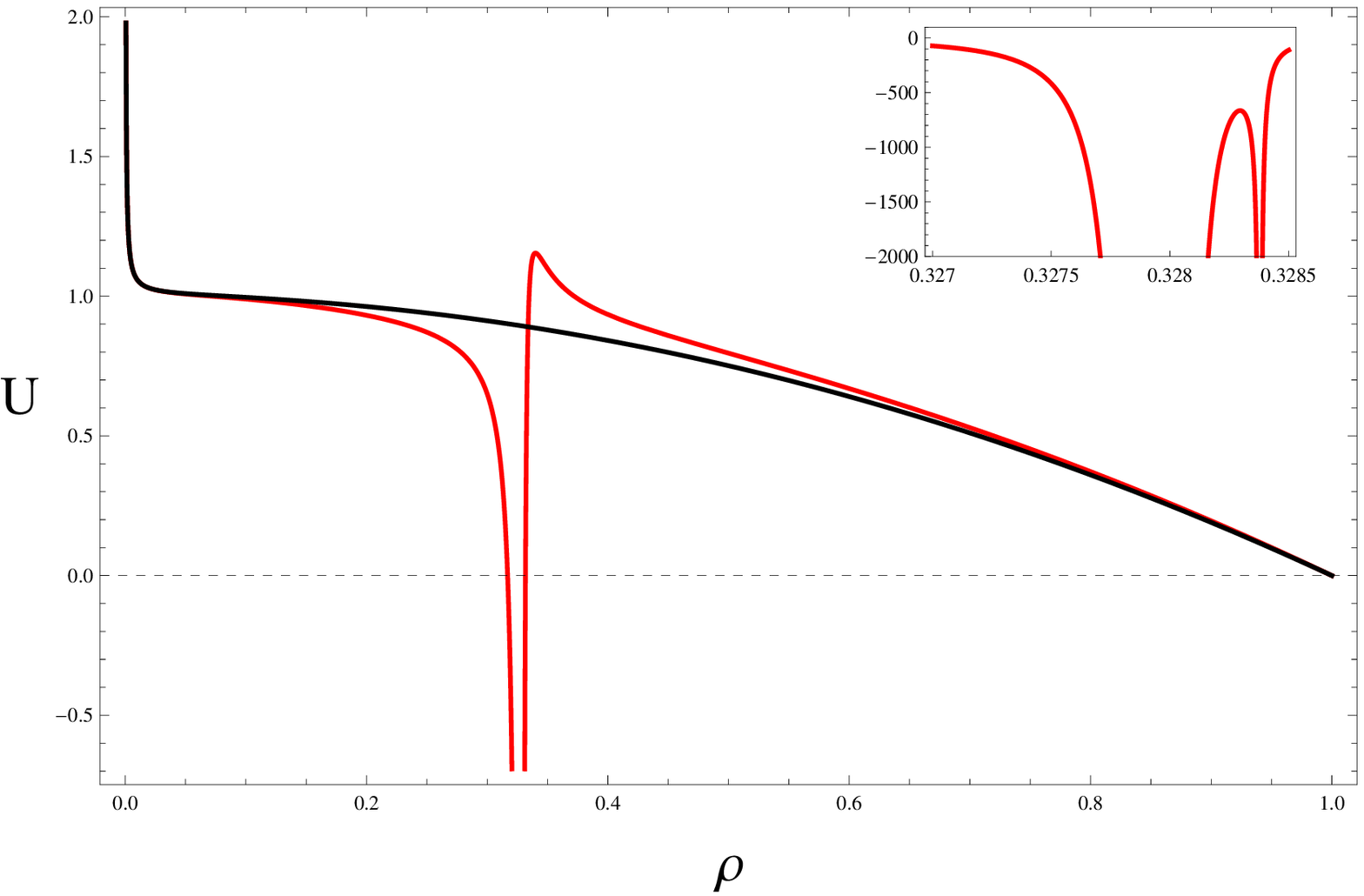}
\caption{Comparison of the effective potential for $\mu<0$ and
$\scrq^2>\scrq_c^2$ (red) with that for Einstein gravity (black). The
red curve is given for $\lambda=-0.1$, $\mu=-0.0003$ and $\scrq=40$.
For these gravitational couplings, the critical momentum is
approximately $\scrq_c\simeq7.27$. The inset shows a close-up of the
effective potential to resolve the two separate singularities at
$\rho_A=0.328378$ and $\rho_D=0.327955$. (Note that in keeping with the
discussion at eq.~\reef{schro99a}, the sign of $U$ in the inset has
been reversed on the interval $\rho_A>\rho>\rho_D$.)}	
\labell{compare-beyond} }

However, the singularities introduce a new boundary surface into the
problem and we argue that this also leads to a additional set of new
unstable modes, as follows: The Einstein potential vanishes at the
horizon and so solving the Schr\"odinger equation \reef{schro99} with a
negative energy yields two independent solutions, one which grows
exponentially (and diverges at the horizon) and another which decays.
Similarly, there are two asymptotic solutions, one which grows and
another which decays as one approaches the AdS boundary. Now the
potential is smooth throughout $0<\rho<1$ if we are considering pure
Einstein gravity. As a result, the solution which decays at the horizon
is precisely that which grows at the asymptotic boundary and vice
versa. Hence one finds no normalizable solutions with a negative
effective energy which agrees with the result that the black hole is
stable in Einstein gravity. However, if we consider quasi-topological
gravity with $\mu<0$ and $|\scrq|>\scrq_c$, while the effective
potential is well approximated by the Einstein potential for most
radii, a dramatic difference arises with the appearance of the
singularities at $\rho=\rho_A$ and $\rho_D$. Now we have the
possibility of matching a solution in the range $\rho>\rho_A$ which
decays at the horizon to a solution in the range $\rho<\rho_D$ which
decays at the asymptotic boundary. Because of the sign of the effective
energy in the Schr\"odinger equation \reef{schro99a} is flipped in the
range $\rho_D<\rho<\rho_A$, the solution would be oscillatory in this
interval and tuning $\alpha^2$ should allow us to match onto the
decaying solutions with the boundary conditions indicated above. Hence
we should also be able to construct an infinite set of negative energy
states which are localized near the interval $\rho_D<\rho<\rho_A$. It
appears that these unstable modes would naturally be regarded as the
progeny of the diverging number of unstable modes which were found to
accumulate in the limit $\scrq^2\rightarrow\scrq^2_c$, in the previous
section. In any event, our conclusion is that generally when $\mu$ is
negative, the black holes in quasitopological gravity have unstable
quasinormal modes with $|\scrq|>\scrq_c$.

To summarize, our analysis in this section indicates that instabilities
appear in the tensor channel throughout the lower half of the space of
couplings with $\mu<0$ and in the narrow sliver shown in figure
\ref{bs} with $\mu>0$. However, we should re-iterate that our analysis
here only represents a preliminary investigation of the potential
plasma instabilities in five-dimensional quasi-topological gravity. A
thorough and detailed analysis of the quasinormal modes is required to
validate the results derived here and to develop a full picture of the
unstable modes.

\section{Discussion} \label{discuss}

In this paper, we have begun an examination of the holographic
properties of quasi-topological gravity. The main new feature of this
toy model is that it allows us to examine CFT's in which the flux
parameter $t_4$ in eq.~\reef{basic} is nonvanishing. In this case, the
dual CFT cannot be supersymmetric \cite{hofmal} and so these new models
allow us to begin exploring holography in a context which is
fundamentally nonsupersymmetric. In this regard, quasi-topological
gravity differs from Lovelock theories which are consistent with
supersymmetry \cite{parn4}. Of course, conformal fixed points are
believed to occur for a wide variety of nonsupersymmetric gauge
theories \cite{erich}. Further one might speculate that extending some
of these to large $N_c$ and strong coupling may generate a holographic
dual close to Einstein gravity \cite{spec}. Hence, in the spirit of
exploring quasi-topological gravity as a toy model, one may gain new
insights into such conformal fixed points.

One aspect of physics which we explored was the hydrodynamic properties
of the CFT plasma. In particular, we calculated the ratio of shear
viscosity to entropy density \reef{etas}. Similar results have already
been found in a framework where the curvature-cubed terms were treated
perturbatively \cite{bd}. Eq.~\reef{etas} should reduce to these
results in the regime of small couplings where one only keeps the terms
linear in $\lambda$ and $\mu$. Our full nonperturbative calculation is
straightforward using the techniques developed in \cite{newwald} and
produces a final expression which contains contributions which are
nonlinear in the gravitational couplings, \eg proportional to
$\mu\,\lambda^2$. This differs from cubic Lovelock theories in higher
dimensions which also contain curvature cubed interactions. In fact, in
this case, $\eta/s$ is independent of the coupling constant controlling
the curvature-cubed interactions \cite{LLrefs,ramy} and remains simply
linear in the curvature squared coupling, as in GB gravity
\cite{shenker1,spain1,GBanyd}. Of course, these theories are also
distinguished from the present case since $t_4$ remains zero in the
Lovelock theories despite the appearance of the curvature cubed
interaction \cite{LLrefs}.

Considering the value of $\eta/s$ when $t_4\ne 0$, we find no dramatic
behaviour. The value smoothly increases or decreases as we move away
from the axis into the $(\lambda,\mu)$ plane, as illustrated in figure
\ref{etaplot}. One point that this analysis makes clear is that
$\eta/s=1/4\pi$ is simply a codimension one surface (here a contour)
which cuts through the space of gravitational couplings or
alternatively through the space of parameters which differentiate the
dual CFT's. The main feature that distinguishes this surface is that it
runs through the point where the bulk theory is Einstein gravity, which
we favour as theorists. This illustrates that even if $\eta/s$ was
found to be precisely $1/4\pi$ for some system arose in nature (\eg the
quark-gluon plasma or a trapped atomic gas), there is no guarantee that
the holographic dual would anywhere be close to a theory of Einstein
gravity.

This holographic model also illustrates the point that the CFT plasma
can readily achieve $\eta/s<1/4\pi$. Of course, even though the
originally conjectured KSS bound on $\eta/s$ has been proven incorrect,
there are still general arguments to suggest that this ratio should
satisfy some lower bound \cite{hydro,bek}. Hence the question naturally
arises as to the precise nature of such a bound. Holographic models
provide an excellent theoretical framework to study this question, as
they provide access to a variety of strongly coupled fluids in this
`KSS regime' where $\eta/s$ is unusually small.

In the present case of holographic fluids modeled by quasi-topological
gravity, this ratio reaches a minimum value at the upper corner of the
allowed parameter space with $\eta/s\simeq(0.4140)/(4\pi)$, as given in
eq.~\reef{min}. It may be of interest to note that at the point, the
dual CFT has $t_2=0$ while $t_4$ reaches its maximal value with
$t_4=15/2$. It is clear that this value does not represent a
fundamental bound. If one explores holographic models with GB gravity
in higher dimensions, one finds the minimum ratio is ${\eta}/{s}
\simeq(0.4139)/(4\pi)$ for $D\simeq 9.207$ \cite{spain1,GBanyd}. This
analysis was also extended to Lovelock gravity with a cubic curvature
interaction for $D\ge7$ \cite{LLrefs}. Initially here, it appeared that
$\eta/s$ could be pushed to zero (or even negative values) but taking
care to account for plasma instabilities, one finds that the
hyrdodynamic results are only reliable down to
${\eta}/{s}\sim(0.3938)/(4\pi)$ \cite{notsofast}. The cubic interaction
of quasi-topological gravity can also be extended to higher $D\ge7$
\cite{quasi} and so it would be interesting to examine the
contributions of such interactions to higher dimensional holographic
models. One might note the Lovelock models introduce additional
parameters to distinguish the dual CFT's but still have $t_4=0$. While
it is perhaps not surprising then, these results explicitly demonstrate
that many more parameters in the CFT will effect the value of $\eta/s$
than simply $\A$, $\B$ and $\C$, the three couplings which fix the the
three-point function of the stress tensor.

While the various holographic models above all seem to point to a
minimum value around ${\eta}/{s}\sim(0.4)/(4\pi)$, it seems clear that
this collection of models only explores a limited parameter space. One
might imagine that one can continue to systematically lower $\eta/s$ by
continuing explore a wider range of CFT's by adding more and more
interesting couplings in the dual gravitational theory (and without
introducing any other pathologies in the holographic model). In fact,
there seems to be some evidence in favour of such a scenario at least
in very high dimensions \cite{reallysmall}.

There is one difference between quasi-topological gravity and GB
gravity (or the more general Lovelock theories) which is particularly
striking. All of the Lovelock gravity models are distinguished by
having second order equations of motion while the general equations in
quasi-topological gravity are fourth order. It seems that the latter is
inevitable in order to produce a holographic theory where $t_4\ne 0$
since the Lovelock theories are in fact the most general gravitational
theories with second order equations of motion \cite{lovel}. To develop
some intuition for such higher order equations, we might establish an
analogy with a higher-derivative scalar field equation (in flat space)
 \be
 \left(\Box + \frac{a}{M^2} \Box^2\right) \phi=0\,.
 \labell{analog}
 \ee
Here we imagine $M^2$ is some high energy scale and $a$ is the
dimensionless coupling that controls the strength of the
higher-derivative term. The (flat space) propagator for this scalar can
be written as
 \be
   \frac{1}{q^2(1- a\, q^2/M^2)} = \frac{1}{q^2} -
   \frac{1}{q^2-M^2/a}\,.
 \labell{propell}
 \ee
Now the $1/q^2$ pole is associated with the regular modes which are
easily excited at low energies. The second pole $1/(q^2-M^2/a)$ is
associated with ghost modes that appear out at the high energy scale.
Depending on the sign of $a$, these new modes may have a regular mass
($a<0$) or be tachyonic ($a>0$). Further writing out the dispersion
relation for the ghost modes, we have
 \be
q^2-M^2/a = -\omega^2 + (k^i)^2 - M^2/a =0\,.
 \labell{propell2}
 \ee
As noted, when $a$ is negative, the mass above has the `right sign' and
these modes only go on-shell when $\omega^2 \sim M^2/|a|$, \ie at very
high energies. On the other hand, if $a$ is positive, the modes are
`unstable' and in this case, we can bring these modes on-shell above a
certain threshold of large spatial momentum, \ie $(k^i)^2 \sim M^2/a$.
Comparing this discussion to our analysis of the tensor channel
equation for quasi-topological gravity, the coupling $\mu$ for the
curvature-cubed interactions would play a role analogous to $a$ above.
In parallel with the present scalar theory in section \ref{bug}, we
found that a new set of unstable modes appears above a certain momentum
threshold for a particular sign of $\mu$, \ie $\scrq^2\ge\scrq_c^2$ for
$\mu<0$.

One important point that arises in the scalar field model is that the
extra high energy modes are ghosts for either sign of $a$, as seen from
the overall sign of their contribution to the propagator
\reef{propell}. Hence a natural worry would be that ghosts must also
appear in quasi-topological gravity but we will argue that this is not
the case, in the following. In section \ref{plastab}, we have found new
unstable modes in quasi-topological theory. This certainly indicates
that working on the uniform black hole background is problematic but it
is not clear that they indicate that there is a fundamental pathology
in the form of ghost modes. One important difference between the
equations in quasi-topological gravity and in the simple scalar model
are that the former are not Lorentz invariant (in the gauge theory
directions). Of course, the lack of Lorentz invariance is not a
surprise since the black hole background is dual to a uniform finite
temperature plasma which defines a preferred reference frame. Hence the
same statement would apply even if we were considering graviton modes
in a black hole solution of Einstein gravity. However, the discrepancy
is more significant for the higher derivative terms here. For example,
while a $\scrq^4$ term appears in eq.~\reef{fulldisp} there is no
$\scrw^4$ contribution. Hence it is not clear whether the additional
instabilities appearing in the black hole background are associated
with ghost modes.

Let us consider this point further. At zero temperature, the background
spacetime reduces to simply AdS$_5$ and as noted before with
eq.~\reef{fullEin}, the linearized equations of motion reduce to the
second order equations of Einstein gravity. Hence in this limit, both
the higher derivative contributions and the additional unstable modes
vanish for any values of $\lambda$ and $\mu$. As noted in \cite{quasi},
the higher derivative terms appear in the linearized equations of
motion through couplings to the background curvature when one is
considering fluctuations around a nontrivial background spacetime. In
particular, these terms arise from a nontrivial Weyl curvature in the
background, \eg for a transverse traceless mode (\ie $\nabla^a h_{a b}
=0$) and $h^a{}_a=0$), the four-derivative terms can be written as
\cite{quasi}:\footnote{Here we adopt the standard notation:
$T_{(ab)}=\frac{1}{2}\left( T_{ab} + T_{ba}\right)$.}
 \beq
C^{ cd ef}\,h_{ de; cf\left( a b\right)} +2\left(\Box h^{
c}{}_{\left(\right. a }\right){}^{\!; de}C_{| cd e| b\left.\right)}
+2\,\Box^2h^{ cd }C_{ c a  d b}+g_{ a  b}\left(\Box h_{ cd }\right)_{;e
f}C^{ ce d f}\,.\labell{fourder}
  \eeq
Hence the specific features of any instabilities will always depend on
the details of the background geometry under study. It may be of
interest to explicitly repeat the analysis of section \ref{plastab} for
other nontrivial configurations, \ie one simple example would be the
confining phase represented by the AdS soliton \cite{gary2}.

Another interesting aspect of these higher derivative terms
\reef{fourder} is that they will vanish in the asymptotic region with
AdS boundary conditions, since the Weyl curvature will vanish there.
Hence any unstable modes associated with these terms will be `confined'
to the interior, \eg near the horizon. Hence in the dual CFT, the
instabilities will be associated with dynamics of infrared excitations
and will be insensitive to the ultraviolet details of the theory. Given
this observation and the previous discussion, it seems then that these
problems cannot represent a fundamental pathology, \ie ghosts, in the
theory. Hence we would conclude that the analogy with the scalar field
in eq.~\reef{analog} is simply deficient in this respect.

Still the higher derivative terms and the resulting instabilities are a
worrying aspect of quasi-topological gravity. In particular, we have
made a preliminary analysis of the shear and sound channel equations of
motion. In this case, it appears that the coefficient analogous $A$ in
eq.~\reef{PreSchro} can again have go to zero but when $\mu$ is
positive. Hence one would be tempted to conclude that instabilities
will now appear for positive $\mu$. Unfortunately, this would mean that
the plasma is only stable with $\mu=0$. This makes the need for a
detailed analysis of the quasinormal modes even more pressing to
develop a full picture of the instabilities.

To close, let us re-iterate that quasi-topological gravity was
introduced as a toy model to study extensions of the usual AdS/CFT
correspondence. We have not identified an approach by which the new
gravitational action \reef{action} emerges from a UV complete theory
and we have no reason to expect that the new theory is radiatively
stable. If one were tempted to construct a full quantum version of
quasi-topological gravity, another issue which would need to be
addressed is the appearance of several different vacua (corresponding
to the roots of eq.~\reef{fin}) and in particular vacua in which the
graviton is a ghost. It is not clear what the role of these vacua would
be in, \eg a path integral formulation of the theory.\footnote{A
related issue was studied for Lovelock gravity in \cite{extra} but
found not to be problematic.} All of these considerations as well as
the appearance of instabilities at large momenta reinforce the idea
that quasi-topological gravity (as well as Lovelock theories of
gravity) should only be treated as toy models which may give us insight
into the long wavelength physics, \eg hydrodynamic behaviour, of
strongly coupled conformal fixed points.

\acknowledgments It is a pleasure to thank Pavel Kovtun, Brandon
Robinson, Dam Son and Andrei Starinets for useful correspondence and
conversations. RCM would also like to thank the KITP and the Weizmann
Institute for hospitality at various stages of this project. Research
at the KITP is supported by the National Science Foundation under Grant
No. PHY05-51164. Research at Perimeter Institute is supported by the
Government of Canada through Industry Canada and by the Province of
Ontario through the Ministry of Research \& Innovation. RCM also
acknowledges support from an NSERC Discovery grant and funding from the
Canadian Institute for Advanced Research. MFP is supported by the
Portuguese Fundacao para a Ciencia e Tecnologia, grant
SFRH/BD/23438/2005. MFP would also like to thank the Perimeter
Institute for hospitality at various stages of this project.

  \end{document}